\documentclass[journal]{IEEEtran}
\IEEEoverridecommandlockouts
\usepackage[utf8]{inputenc}
\usepackage{cite}
\usepackage{amsmath,amssymb,amsfonts}
\usepackage{algorithmic}
\usepackage{graphicx}
\usepackage{textcomp}
\usepackage{xcolor}
    
\usepackage{regensky}
\usepackage{hyperref}

\begin{document}

\markboth{\parbox[t]{20cm}{This work has been submitted to the IEEE for possible publication.\newline Copyright may be transferred without notice, after which this version may no longer be accessible.}}{}

\title{Motion-Plane-Adaptive Inter Prediction\\in 360-Degree Video Coding}

\author{Andy Regensky, Christian Herglotz, \textit{Member, IEEE}, and André Kaup, \textit{Fellow, IEEE}%
\thanks{A. Regensky, C. Herglotz, and A. Kaup are with the Chair of Multimedia Communications and Signal Processing, Friedrich-Alexander University Erlangen-Nürnberg, Cauerstr. 7, 91058 Erlangen, Germany. E-mail: andy.regensky@fau.de, christian.herglotz@fau.de, andre.kaup@fau.de.}%
\thanks{The authors gratefully acknowledge that this work has been supported by the German Research Foundation (DFG) under project number 418866191.}
}

\maketitle

\begin{abstract}
  Inter prediction is one of the key technologies enabling the high compression efficiency of modern video coding standards.
  360-degree video needs to be mapped to the 2D image plane prior to coding in order to allow compression using existing video coding standards.
  The distortions that inevitably occur when mapping spherical data onto the 2D image plane, however, impair the performance of classical inter prediction techniques.\newline
  In this paper, we propose a motion-plane-adaptive inter prediction technique (MPA) for 360-degree video that takes the spherical characteristics of 360-degree video into account.
  Based on the known projection format of the video, MPA allows to perform inter prediction on different motion planes in 3D space instead of having to work on the - in theory arbitrarily mapped - 2D image representation directly.
  We furthermore derive a motion-plane-adaptive motion vector prediction technique (MPA-MVP) that allows to translate motion information between different motion planes and motion models.
  Our proposed integration of MPA together with MPA-MVP into the state-of-the-art H.266/VVC video coding standard shows significant Bjøntegaard Delta rate savings of 1.72\% with a peak of 3.97\% based on PSNR and 1.56\% with a peak of 3.40\% based on WS-PSNR compared to the VTM-14.2 baseline on average.
\end{abstract}


\section{Introduction}\label{sec:introduction}

\IEEEPARstart{T}{he} all-around field of view of 360-degree video combined with stereo capture technologies capable of creating a sense of virtual depth allows to create virtual reality (VR) experiences providing an unprecedented feeling of immersion.
With the increasing availability of affordable VR headsets, 360-degree video has the potential to be the next major evolutionary step in video technology.

\begin{figure}[t]
  \centering
  \subfloat[ERP-projected 360-degree image\label{fig:360image:erp}]{%
  \includegraphics[width=\linewidth]{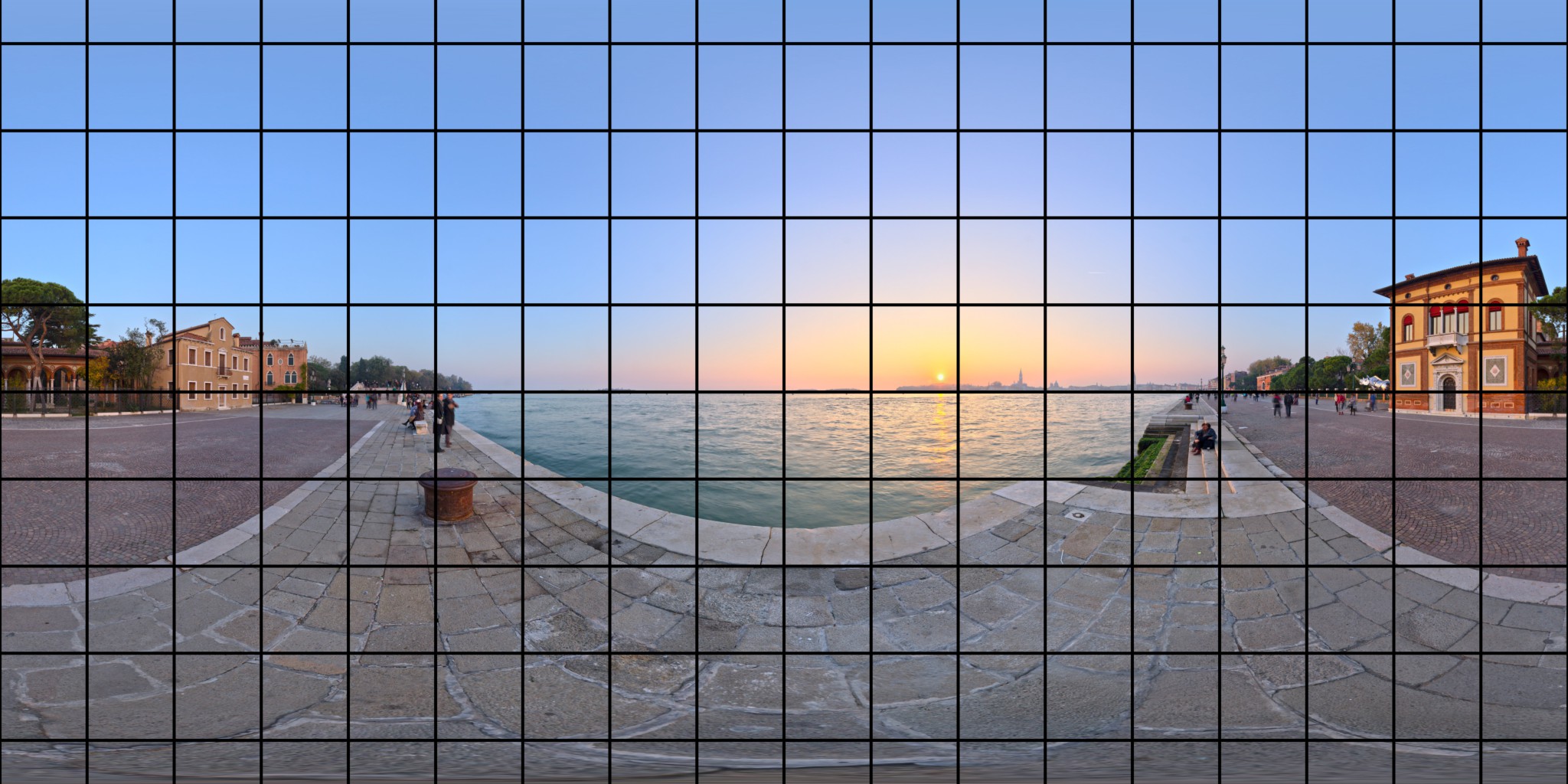}}
  \\
  \subfloat[sphere mapping (front)\label{fig:360image:front}]{%
  \includegraphics[width=0.45\linewidth]{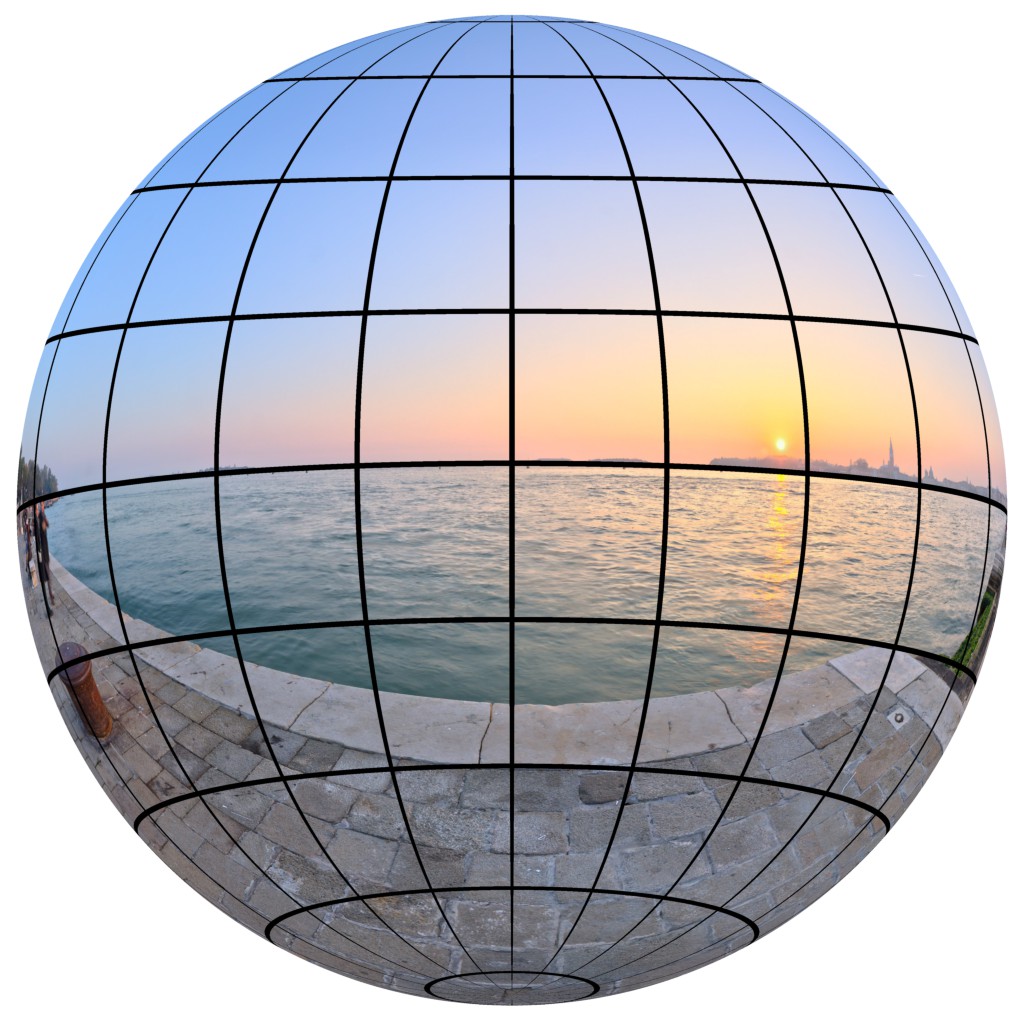}}
  \hfill
  \subfloat[sphere mapping (back)\label{fig:360image:back}]{%
  \includegraphics[width=0.45\linewidth]{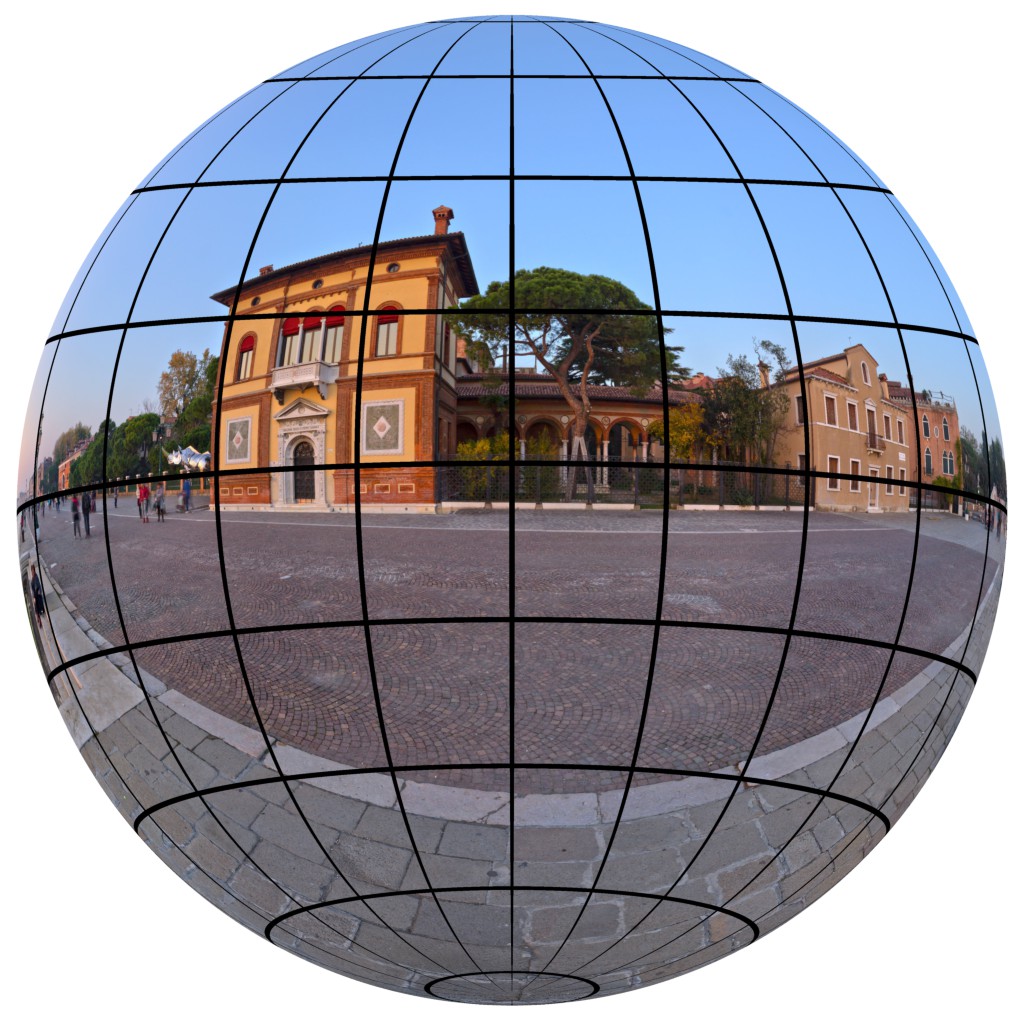}}

  \caption{(a) A 360-degree image mapped to the 2D image plane using an equirectangular projection (ERP). (b), (c) The 360-degree image mapped to the unit sphere in 3D space visualizing (b) the front view and (c) the back view.}
  \label{fig:360image}
\end{figure}

The delivery of immersive 360-degree VR experiences can be divided into three main areas: capture, storage \& distribution, and display.
Our research addresses the second area, more specifically, the development and design of improved coding techniques for 360-degree video compression.
Typically, a 2D representation of the 360-degree video is required to allow compression using existing video coding techniques such as the H.264/AVC~\cite{Wiegand2003}, the H.265/HEVC~\cite{Sullivan2012} or the H.266/VVC~\cite{Bross2021, Bross2021a} video coding standards.
Fig.~\ref{fig:360image:erp} shows an example of a 360-degree image mapped to the 2D image plane through an equirectangular projection (ERP).
While this is one of the most common 360-degree projection formats, there exist a plethora of other formats including various variations of cubemap projections, segmented and rotated sphere projections, or octa- and icosahedron projections to name only a few~\cite{Ye2020a}.

Compared to coding conventional perspective video, the compression performance of modern video codecs is notably reduced for non-perspective projection formats~\cite{Wien2017, Bross2021a}.
One reason for this are the distortions that inevitably occur due to the mapping of the spherical 360-degree video to the 2D image plane.
Fig.~\ref{fig:360image:front} and~(c) show the spherical domain representation of the ERP-projected 360-degree image in Fig.~\ref{fig:360image:erp}.
While the black lines of constant azimuthal (vertical) and polar (horizontal) angles form a block structure in the ERP-projected image, this is not the case in the spherical domain.
It is clearly visible that the different blocks become increasingly distorted with a higher distance to the equator.

This makes the investigation of inter prediction techniques in the context of 360-degree video coding especially important.
In classical translational inter prediction, each block in a regarded frame is predicted from a shifted block in a known reference frame whereby the block retains its size and shape.
However, as can be seen in Fig.~\ref{fig:360image}, the further to the poles an object lies, the higher its distortion in the ERP-representation.
The classical translational inter prediction technique is not able to replicate the complex block distortions in the ERP-domain that result from camera and object motion in any capacity.
The design and optimization of coding techniques taking the known characteristics of 360-degree video into account is therefore vital to support the development towards immersive virtual realities.
Motivated by this, different approaches have been tested including the design of improved projection formats~\cite{Ye2020a}, the introduction of new coding tools~\cite{Ye2020}, and the development of adaptive streaming techniques~\cite{Hosseini2016, DelaFuente2019}.

In this work, we propose a novel \textit{motion-plane-adaptive} inter prediction technique (MPA) for 360-degree video that allows to perform inter prediction on different motion planes in 3D space.
Any motion on these planes is modeled purely using horizontal and vertical shifts while the motion planes themselves can be oriented freely in 3D space.
In this way, MPA takes both the spherical characteristics of 360-degree video and the translational nature of most camera and object motion into account.
MPA thus is able to more accurately reproduce the resulting pixel shifts in the 2D projection domain than classical translational techniques are able to.
Due to their narrow field of view, such 3D space considerations are not necessary for conventional perspective video.
To further improve the performance of MPA and make it compatible to existing inter prediction techniques, we additionally derive an efficient method to transfer motion information between different motion planes and motion models.

The remainder of the paper is organized as follows.
In Section~\ref{sec:related}, the traditional inter prediction procedure is briefly recapitulated and an overview over related approaches to improving 360-degree video coding is given.
The main contributions of this work are compiled in Section~\ref{sec:mpa-inter}.
It introduces the projection functions that are required for motion-plane adaptivity including a generalized formulation of the perspective projection, presents the motion-plane-adaptive motion model, derives an adapted motion vector prediction method, and explains the integration of the proposed concepts into the H.266/VVC video coding standard.
Section~\ref{sec:performance} then evaluates the performance of MPA presenting both numerical and visual results.
Finally, Section~\ref{sec:conclusion} concludes the paper.

\section{Background and Related Work}\label{sec:related}

Inter prediction is a crucial component of any modern hybrid video codec, where the term hybrid refers to a combination of predictive and transform-based coding.
Typically, the current frame \mbox{$\vec{I}_\text{cur} \in \mathbb{R}^{U\times V}$} of size $U \times V$ pixels to be coded is subdivided into individual blocks $\vec{B}_\text{cur} \in \mathbb{R}^{M\times N}$ of size $M \times N$ pixels and each block is coded individually.
In a first step, a prediction $\vec{B}_\text{pred} \in \mathbb{R}^{M\times N}$ is formed for each block based on its causal spatial and temporal neighborhood.
Most video codecs allow either intra (spatial) or inter (temporal) prediction for a given block.

In a second step, the residual signal $\vec{B}_\text{res} \in \mathbb{R}^{M\times N}$ between the predicted block $\vec{B}_\text{pred}$ and the actual block $\vec{B}_\text{cur}$
\begin{align}
  \vec{B}_\text{res} = \vec{B}_\text{cur} - \vec{B}_\text{pred}.
\end{align}
is converted to a transform domain and the resulting signal is quantized and entropy coded~\cite{Wiegand2003, Sullivan2012, Bross2021a}.

At the decoder, the prediction is formed analog to the prediction procedure at the encoder, before the decoded residual is added to the prediction yielding the reconstructed block~\mbox{$\widehat{\vec{B}}_\text{cur} \in \mathbb{R}^{M\times N}$}.
To ensure that both the encoder and the decoder are able to arrive at the same prediction, additional side information is signaled to control the prediction procedure.
As such, a flag indicates whether intra or inter prediction is used and further control mechanisms specify additional prediction information.
In traditional inter prediction, this prediction information includes motion information that is shared by all pixels within the block.
The precise motion information that needs to be signaled depends on the applied motion model.

Using the translational motion model
\begin{align}
  \vec{m}_\text{t}(\vec{p}, \vec{t}) = \vec{p} + \vec{t},\label{eq:translational-motion-model}
\end{align}
the motion is described by the motion vector \mbox{$\vec{t} = (\Delta u, \Delta v)^T \in \mathbb{R}^2$} that shifts the pixel coordinate \mbox{$\vec{p} = (u, v)^T \in \mathbb{R}^2$} by $\Delta u$ pixels in horizontal $u$-direction and $\Delta v$ pixels in vertical $v$-direction.

In a video codec, the encoder is now responsible for searching the best matching motion vector $\vec{t}^*$ for each block that leads to the best possible quality at the lowest possible rate in a process called rate-distortion optimization~\cite{Sullivan1998}.
A motion compensated or predicted image $\vec{I}_\text{pred} \in \mathbb{R}^{U\times V}$ can then be obtained both at the encoder and at the decoder by extracting the motion compensated pixel values from the reference image as
\begin{align}
  \vec{I}_\text{pred}(\vec{p}) = \vec{I}_\text{ref}(\vec{m}_\text{t}(\vec{p}, \vec{t}^*))~~\forall~\vec{p} \in \mathcal{B} \label{eq:tmc}
\end{align}
for all blocks in the image, where $\mathcal{B}$ denotes the set of pixel coordinates within the regarded block $\vec{B}_\text{pred}$, and $\vec{I}_\text{ref} \in \mathbb{R}^{U\times V}$ describes the reference image.
Thereby, $\vec{I}(\vec{p})$ yields the pixel value of image~$\vec{I}$ at pixel coordinate $\vec{p}$.
Internally, a suitable interpolation method is required in order to access pixel values at fractional pixel positions.

To improve the compression efficiency of video codecs for 360-degree video, a lot of effort has gone into designing improved motion models based on the 360-degree video's spherical domain representation.
Please note that similar to motion models available for traditional 2D video coding, the primary focus is not to describe the true underlying 3D object motion but to achieve highly compressible motion representations suitable for efficient inter prediction.

In~\cite{Li2017, Li2019}, \textit{Li et al.} propose a 3D translational motion model, where all pixels in a regarded block on the sphere are shifted in 3D space according to a 3D motion vector derived from the original 2D motion vector.
A similar approach is followed by \textit{Wang et al.} in~\cite{Wang2017, Wang2019}, where the 3D motion vector is derived based on the assumption that two neighboring blocks adhere to the same motion in 3D space.
In~\cite{Vishwanath2017, Vishwanath2018a}, \textit{Vishwanath et al.} introduce a rotational motion model, where all pixels in a regarded block are rotated on the sphere according to a 2D motion vector that is interpreted as rotation angles.
In~\cite{Vishwanath2018}, they furthermore propose a motion model, where all pixels in a regarded block are rotated along geodesics that are oriented along a known global camera motion vector to better model translational camera motion.
In~\cite{DeSimone2017}, \textit{De Simone et al.} propose a motion model, where the motion is applied on a plane tangential to the regarded block center in the spherical domain representation using a gnomonic projection.
In~\cite{Marie2021}, \textit{Marie et al.} propose a combination of the rotational and tangential motion models, where an optional third motion parameter allows to vary the depth of the pixels in 3D space.

Besides the development of new motion models, researchers have also investigated other approaches to achieve an improved compression efficiency for 360-degree video.
In~\cite{Sauer2017}, \textit{Sauer et al.} propose an improved reference frame padding technique for 360-degree projections where the spherical video is projected to polytopes such as the cubemap projection.
They exploit the known face geometry and pad each face with the available pixel data from connected faces.
In~\cite{He2017}, \textit{He et al.} build upon this technique and propose an improved padding for the equirectangular projection that performs a geometrically correct wrap-around padding instead of the usual replicate padding.
Due to its low computational complexity, this technique is integrated into the novel H.266/VVC video coding standard.

In~\cite{Sauer2018}, \textit{Sauer et al.} furthermore propose an adapted deblocking filter that takes the spherical geometry of the 360-degree video into account to reduce artifacts.
A related, but considerably less complex technique is integrated into the H.266/VVC video coding standard that disables loop filters across explicitly signaled virtual boundaries~\cite{Bross2021a}.

In~\cite{Herglotz2019}, \textit{Herglotz et al.} propose to ignore inactive regions in the projection formats during rate-distortion optimization, residual transformation and in-loop filtering.
Through this concept, they are able to reach notable rate savings in unpacked projection formats which contain areas that are irrelevant for the spherical representation of the 360-degree video.

In~\cite{Zhou2020}, \textit{Zhou et al.} propose an improvement to the rate-distortion-optimization that takes the sampling density of the equirectangular and cubemap projections into account.
They derive appropriate quantization parameters on block level based on the spherical sampling density of the applied 360-degree projection

\section{Motion-Plane-Adaptive Inter Prediction}\label{sec:mpa-inter}

At the core of MPA lies the eponymous motion-plane-adaptive motion model that allows to model motion on arbitrary motion planes in 3D space.
We originally proposed a motion compensation technique capable of modeling motion on different motion planes in the context of fisheye video in~\cite{Regensky2021}.
In this paper, we build upon this concept to allow inter prediction for 360-degree video on different motion planes in 3D space.
The application to inter prediction turned out to require a number of additional considerations and adaptions in order to achieve competitive compression results.

This section describes the final outcome of our investigations and is subdivided into multiple subsections to present MPA in its entirety.
Subsection A introduces the mathematical basics of 360-degree projections and proposes a generalized perspective projection capable of representing 360-degree image and video despite its physical limitations.
Based on these projections, Subsection B then describes the proposed motion-plane-adaptive motion model in detail.
Subsection C derives an adapted motion vector prediction that is capable of translating motion information between different motion planes and motion models for use in motion vector prediction.
Finally, Subsection D demonstrates the broad applicability of the proposed MPA technique by tightly integrating it into the state-of-the-art H.266/VVC video coding standard taking all dependent tools in the coding chain into account.

\subsection{Projections}\label{subsec:projections}

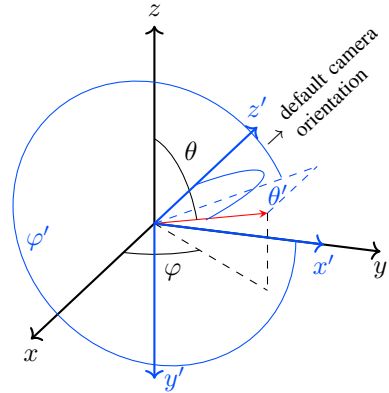
\begin{figure}[t]
  \centering
  \tdplotsetmaincoords{70}{110} 
\begin{tikzpicture}[scale=4,tdplot_main_coords]

\coordinate (O) at (0,0,0);

\draw[thick,->] (O) -- (1.2,0,0) node[anchor=north]{$x$};
\draw[thick,->] (O) -- (0,0.8,0) node[anchor=north]{$y$};
\draw[thick,->] (O) -- (0,0,0.7) node[anchor=south]{$z$};

\pgfmathsetmacro{\rvec}{.8}
\pgfmathsetmacro{\thetavec}{70}
\pgfmathsetmacro{\phivec}{50}
\tdplotsetcoord{P}{\rvec}{\thetavec}{\phivec}
\draw[-stealth,color=red] (O) -- (P);
\draw[dashed, color=black] (O) -- (Pxy);
\draw[dashed, color=black] (P) -- (Pxy);
\tdplotdrawarc{(O)}{0.3}{0}{\phivec}{anchor=north,xshift=4}{$\varphi$}
\tdplotsetthetaplanecoords{\phivec}
\tdplotdrawarc[tdplot_rotated_coords]{(0,0,0)}{0.3}{0}{\thetavec}{anchor=south west, xshift=-2}{$\theta$}

\tdplotsetrotatedcoords{0}{-90}{90}  
\draw[thick,color=blue!70!cyan,tdplot_rotated_coords,->] (O) -- (.6,0,0) node[anchor=north]{$x^\prime$};
\draw[thick,color=blue!70!cyan,tdplot_rotated_coords,->] (O) -- (0,.55,0) node[anchor=west]{$y^\prime$};
\draw[thick,color=blue!70!cyan,tdplot_rotated_coords,->] (O) -- (0,0,1) node[anchor=south](optical_axis){$z^\prime$};
\node[below left = -44pt and -50pt of optical_axis, rotate=43, align=left] {\footnotesize \textrightarrow\ \parbox{6em}{default camera orientation}};

\draw[dashed, color=blue!70!cyan] (O) -- (Pyz);
\draw[dashed, color=blue!70!cyan] (P) -- (Pyz);
\tdplotdrawarc[tdplot_rotated_coords,color=blue!70!cyan]{(O)}{0.5}{0}{334}{anchor=west}{$\varphi^\prime$};
\tdplotsetrotatedthetaplanecoords{334};
\tdplotdrawarc[tdplot_rotated_coords,color=blue!70!cyan]{(O)}{0.4}{0}{130}{anchor=west,yshift=-8,xshift=-2,color=blue!70!cyan}{$\theta^\prime$};

\end{tikzpicture}
  \vspace*{-1em}
  \caption{The employed 3D coordinate systems. The black coordinate system $(x, y, z)$ describes the main system orientation where $y$ is oriented horizontally, $z$ is oriented vertically, and $x$ is oriented perpendicular to $y$ and $z$. The default camera is positioned at the origin and oriented in negative $x$-direction. $\theta$ and $\varphi$ denote the corresponding polar and angular angles in spherical coordinates. The blue coordinate system $(x^\prime, y^\prime, z^\prime)$ describes an intermediate system used for the perspective projection where the virtual perspective camera is oriented in positive $z^\prime$-direction. The corresponding polar and angular angles $\theta^\prime$ and $\varphi^\prime$ are given in blue.}
  \label{fig:3d-coordinate-systems}
\end{figure}

As MPA is based on the known mappings between the 2D image plane and the 3D space representations of a 360-degree video, a general formulation of these mappings in the form of projection functions is required.
Any valid projection function~$\boldsymbol{\xi} : \mathcal{S} \rightarrow \mathbb{R}^2$ is invertible and describes the relation between a 3D space coordinate $\vec{s} = (x, y, z)^T \in \mathcal{S}$ on the unit sphere and the corresponding pixel coordinate $\vec{p} = (u, v)^T \in \mathbb{R}^2$ on the 2D image plane, where $\mathcal{S} = \{ \vec{s} \in \mathbb{R}^3\ |\ \lVert\vec{s}\rVert_2 = 1 \}$ describes the set of all coordinates on the unit sphere.
The inverse projection function $\boldsymbol{\xi}^{-1} : \mathbb{R}^2 \rightarrow \mathcal{S}$ maps the 2D image plane coordinate back to the unit sphere in 3D space.

The orientation of the applied 3D coordinate sytem $(x, y, z)$ is visualized in black in Fig.~\ref{fig:3d-coordinate-systems}.
Thereby, $y$ is oriented horizontally, $z$ is oriented vertically and $x$ is oriented perpendicular to $y$ and $z$.
The default camera is positioned at the origin and oriented in negative $x$-direction.
The rotated blue coordinate system $(x^\prime, y^\prime, z^\prime)$ is an intermediate system for the generalized perspective projection, which will be introduced later.

For MPA, two projection functions are important.
First, the employed 360-degree projection $\boldsymbol{\xi}_\text{o}$ of the given video, and second, the perspective projection $\boldsymbol{\xi}_\text{p}$ for representing motion on the desired motion planes in 3D space.

\subsubsection*{Equirectangular Projection}
The equirectangular projection $\boldsymbol{\xi}_\text{erp}$ is a popular and widely applied example of a general 360-degree projection.
It maps the polar angle $\theta \in [0, \pi]$ to the vertical $v$-axis and the azimuthal angle $\varphi \in [0, 2\pi]$ to the horizontal $u$-axis of the 2D image plane.
For projecting a 3D space coordinate $\vec{s}$ on the unit sphere to the 2D image plane, its spherical angles $(\theta, \varphi)$ according to Fig.~\ref{fig:3d-coordinate-systems} need to be obtained first through
\begin{align}
  \theta &= \arccos(z), \label{eq:erp_forward:begin} \\
  \varphi &= \arctantwo(y, x),
\end{align}
where $\arctantwo$ describes the four-quadrant arctangent.
The spherical angles are then projected to the 2D image plane yielding the pixel coordinate $\vec{p}_\text{erp} = (u_\text{erp}, v_\text{erp})^T \in \mathbb{R}^2$ as
\begin{align}
  u_{\text{erp}} &= \frac{\varphi}{2\pi} \cdot U, \\
  v_{\text{erp}} &= \frac{\theta}{\pi} \cdot V, \label{eq:erp_forward:end}
\end{align}
where $U$ describes the width and $V$ the height of the 2D image plane in pixels.
Typically, $U = 2V$ in case of the equirectangular projection as the azimuthal angle $\varphi$ has twice the angular range compared to the polar angle $\theta$.
The equirectangular projection function $\boldsymbol{\xi}_\text{erp}$ combines steps~\eqref{eq:erp_forward:begin}-\eqref{eq:erp_forward:end} in a concise expression.

For the inverse equirectangular projection $\boldsymbol{\xi}_\text{erp}^{-1}$, the pixel coordinate on the 2D image plane is projected back to the spherical domain through
\begin{align}
  \varphi &= \frac{u_\text{erp}}{U} \cdot 2\pi, \\
  \theta &= \frac{v_\text{erp}}{V} \cdot \pi,
\end{align}
before the final pixel coordinate on the unit sphere is obtained as
\begin{align}
  x &= \sin(\theta) \cos(\varphi), \\
  y &= \sin(\theta) \sin(\varphi), \\
  z &= \cos(\theta).
\end{align}

\subsubsection*{Generalized Perspective Projection}
Other than most 360-degree projections, the perspective projection $\boldsymbol{\xi}_\text{p}$ represents a real-world physical model of a light ray passing through a pinhole on its way to the 2D image plane, the so-called pinhole model.
To alleviate the problem of low light intensity, perspective lenses have been designed that replicate the general behavior of the pinhole model while also capturing more light.

\begin{figure}[t]
  \centering
  \begin{tikzpicture}[scale=4]
    \small
    \definecolor{better_yellow}{HTML}{F6BE00}
    \pgfmathsetmacro{\rayalphaa}{1.4}
    \pgfmathsetmacro{\rayalphab}{1.5}

    \coordinate (O) at (0, 0);

    \path (O) +(0, -0.6) coordinate (rip_O);
    \path (rip_O) +(-0.6, -0.25) coordinate (rip_bl);
    \path (rip_bl) +(1.0, 0) coordinate (rip_br);
    \path (rip_O) +(0.6, 0.25) coordinate (rip_tr);
    \path (rip_tr) +(-1.0, 0) coordinate (rip_tl);
    \path (rip_O) +(-0.36, -0.1) coordinate (rip_intersection);
    \path (O) +($(rip_intersection) - \rayalphaa*(rip_intersection)$) coordinate (rip_ray_start);

    \path (O) +(0, 0.6) coordinate (vip_O);
    \path (vip_O) +(-0.6, -0.25) coordinate (vip_bl);
    \path (vip_bl) +(1.0, 0) coordinate (vip_br);
    \path (vip_O) +(0.6, 0.25) coordinate (vip_tr);
    \path (vip_tr) +(-1.0, 0) coordinate (vip_tl);
    \path (vip_O) +(-0.36, -0.1) coordinate (vip_intersection);
    \path (O) +($(vip_intersection) - \rayalphab*(vip_intersection)$) coordinate (vip_ray_start);

    \draw[thick] (rip_bl) -- (rip_br) coordinate[midway] (rip_edge_bottom);
    \draw[thick] (rip_br) -- (rip_tr) coordinate[midway] (rip_edge_right);
    \draw[thick] (rip_tr) -- (rip_tl);
    \draw[thick] (rip_tl) -- (rip_bl) coordinate[midway] (rip_edge_left);
    \path (rip_O) -- (rip_edge_right) coordinate[midway] (rip_x);
    \path (rip_O) -- (rip_edge_bottom) coordinate[midway] (rip_y);
    \draw[thick,->] (rip_O) -- (rip_x) node[anchor=west] {$u$};
    \draw[thick,->] (rip_O) -- (rip_y) node[anchor=west] {$v$};

    \path (rip_O |- rip_bl)+(0, -0.1) coordinate (optical_axis_bottom);
    \draw[dashed] (rip_O |- rip_bl) -- (optical_axis_bottom) node[anchor=north] {Optical axis};
    \draw[dashed] (rip_O) -- (O);
    \node[ellipse, fill=carolinablue, opacity=0.5, minimum width=128pt, minimum height=32pt] (lens) at (O) {};
    \node[anchor=east, xshift=-4pt] at (lens.west) {Lens};
    \draw[dashed] (O) -- (vip_O |- vip_bl);

    \draw[thick, better_yellow] (rip_ray_start) -- (rip_intersection);
    \draw[thick, dashed, better_yellow] (vip_ray_start) -- (vip_intersection);
    \node[anchor=west, xshift=2pt] at (rip_ray_start) {Ray 0};
    \node[anchor=west, xshift=2pt] at (vip_ray_start) {Ray 1};

    \fill[fill=white] (vip_bl)--(vip_br)--(vip_tr)--(vip_tl)--cycle;
    \draw[thick] (vip_bl) -- (vip_br) coordinate[midway] (vip_edge_bottom);
    \draw[thick] (vip_br) -- (vip_tr) coordinate[midway] (vip_edge_right);
    \draw[thick] (vip_tr) -- (vip_tl);
    \draw[thick] (vip_tl) -- (vip_bl) coordinate[midway] (vip_edge_left);
    \path (vip_O) -- (vip_edge_right) coordinate[midway] (vip_x);
    \path (vip_O) -- (vip_edge_bottom) coordinate[midway] (vip_y);
    \draw[thick,->] (vip_O) -- (vip_x) node[anchor=west] {$u$};
    \draw[thick,->] (vip_O) -- (vip_y) node[anchor=west] {$v$};

    \node[anchor=east, xshift=-4pt, align=right] at (rip_edge_left.center -| lens.west) {Real\\image\\plane};
    \node[anchor=east, xshift=-4pt, align=right] at (vip_edge_left.center -| lens.west) {Virtual\\image\\plane};

    \path (vip_O |- vip_tl)+(0, 0.1) coordinate (optical_axis_top);
    \draw[dashed] (vip_O) -- (optical_axis_top);

    \filldraw[black] (rip_intersection) circle (0.01);
    \draw[thin, dashed] (rip_O) -- (rip_intersection) node[midway, anchor=south] (rip_r) {\scriptsize$r_{\text{p}0}$};
    \draw pic[draw=black, angle radius=9pt, dashed] {angle=rip_intersection--rip_O--rip_x};
    \node[below right = 6pt and 6pt of rip_O, anchor=west] (rip_phi) {\scriptsize$\varphi_{0}^\prime$};
    \draw pic[draw=black, angle radius=24pt] {angle=rip_ray_start--O--vip_O};
    \node[above right= 11pt and 4.2pt of O, anchor=south] (rip_theta) {\scriptsize$\theta_0^\prime$};

    \filldraw[black] (vip_intersection) circle (0.01);
    \draw[thin, dashed] (vip_O) -- (vip_intersection) node[midway, anchor=south] (vip_r) {\scriptsize$r_{\text{p}1}$};
    \draw pic[draw=black, angle radius=9pt, dashed] {angle=vip_intersection--vip_O--vip_x};
    \node[below right = 6pt and 6pt of vip_O, anchor=west] (vip_phi) {\scriptsize$\varphi_{1}^\prime$};
    \draw pic[draw=black, angle radius=30pt] {angle=vip_ray_start--O--vip_O};
    \node[right = 32pt of O, anchor=east] (vip_theta) {\scriptsize$\theta_{1}^\prime$};

    \path (rip_tr) +(0.05, -0.25) coordinate (f_b);
    \path (f_b) +(0, 0.6) coordinate (f_t);
    \path (f_t) +(0, 0.6) coordinate (f_tt);
    \draw[|-|] (f_b) -- (f_t) node[midway, anchor=west] (focal_length) {$f$};
    \draw[-|] (f_t) -- (f_tt) node[midway, anchor=west] (focal_length) {$f$};
\end{tikzpicture}
  \vspace*{-1em}
  \caption{Perspective image planes. Light rays with incident angles $\theta^\prime < \pi/2$ are projected to the real image plane, while light rays with incident angles $\theta^\prime > \pi/2$ are projected to the virtual image plane.}
  \label{fig:perspective-image-planes}
\end{figure}
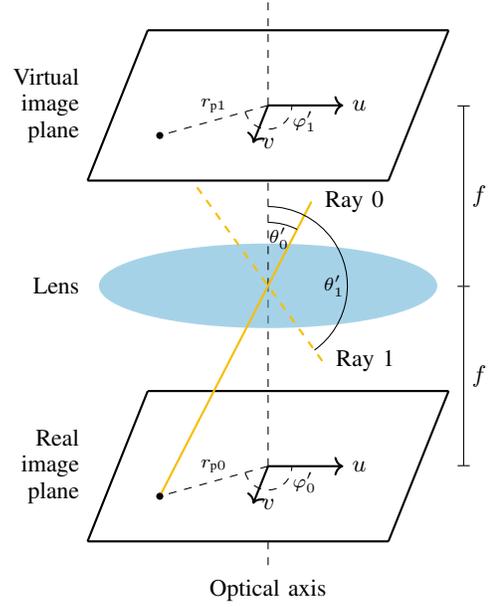

\begin{figure*}[t]
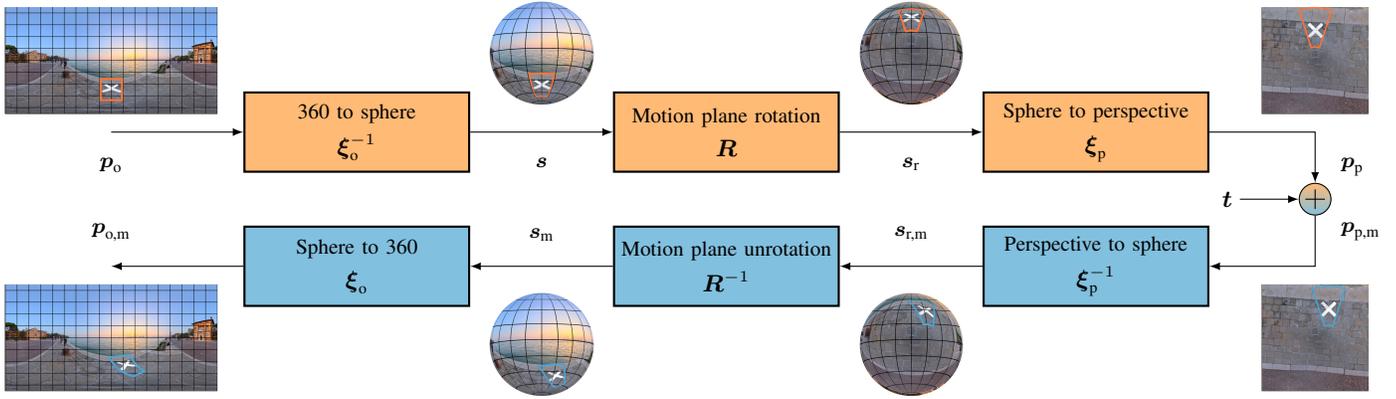

  \centering
  \include{tikz/mpa-schematic}
  \vspace*{-1em}
  \caption{Schematic representation of the motion-plane-adaptive motion model with a visualization of the procedure for an exemplary block in a 360-degree image employing an ERP projection. For visualization, block motion for an exemplary motion vector and rotation matrix is shown. The original block is shown in orange in the top row and the corresponding moved block is shown in blue in the bottom row. The applied motion plane rotation matrix rotates the motion plane by $\pi/2$ around the $y$-axis. It is clearly visible that the distortions in the equirectangular domain resulting from translational motion on the street surface are accurately replicated by the motion-plane-adaptive motion model.}
  \label{fig:mpa-schematic}
\end{figure*}

The general behavior of the perspective projection is visualized in Fig.~\ref{fig:perspective-image-planes}, where the light rays pass through the optical axis of the lens in a straight line.
The exact position $\vec{p}_\text{p} = (u_\text{p}, v_\text{p})^T \in \mathbb{R}^2$, where a light ray intersects the image plane, is solely defined by its incident angle $\theta^{\prime}$ with respect to the optical axis of the lens, its azimuthal angle $\varphi^{\prime}$, and the focal length $f$ given in pixel units, i.e., the distance between the lens and the image plane.

The incident angle $\theta^{\prime}$ and the azimuthal angle $\varphi^{\prime}$ are related to the pixel coordinate $\vec{s}$ on the unit sphere through the rotated coordinate system $(x^\prime, y^\prime, z^\prime)$ as shown in Fig.~\ref{fig:3d-coordinate-systems}.
Using this relation, the incident angle $\theta^{\prime}$ and the azimuthal angle $\varphi^{\prime}$ are obtained from the pixel coordinate $\vec{s}$ as
\begin{align}
  \theta^{\prime} &= \arccos( -x), \label{eq:perspective:begin} \\
  \varphi^{\prime} &= \arctantwo(-z, y).
\end{align}

As the perspective projection is in general only defined for incident angles $\theta_\text{s}^\prime < \pi/2$, an adjusted procedure for incident angles $\theta_\text{s}^\prime > \pi/2$ is required.
We generalize the perspective projection for incident angles $\theta^\prime > \pi/2$ by projecting the corresponding light rays to a so-called virtual image plane on the opposite side of the lens as visualized in Fig.~\ref{fig:perspective-image-planes}.
A binary variable $b_\text{vip} \in \{0, 1\}$ describes whether the light ray intersects the virtual image plane and is defined based on the light ray's incident angle $\theta^\prime$ as
\begin{align}
  b_{\text{vip}} &= \begin{cases}
                      0 & \text{if $0 \leq \theta^{\prime} < \pi/2$,} \\
                      1 & \text{if $\pi/2 < \theta^{\prime} \leq \pi$.}
  \end{cases}\label{eq:perspective:bvip}
\end{align}

The intersection radius of the light ray with the real or virtual image plane relative to the optical axis of the lens is then calculated based on $b_\text{vip}$ through
\begin{align}
  r_{\text{p}} = \begin{cases}
                   f\tan(\theta^\prime) & \text{if $b_{\text{vip}} = 0$,} \\
                   f\tan(\pi - \theta^\prime) & \text{if $b_{\text{vip}} = 1$.}
  \end{cases}\label{eq:perspective:radius}
\end{align}
Finally, the pixel coordinate $\vec{p}_\text{p}$ of the light ray on the perspective image plane is obtained using
\begin{align}
  u_{\text{p}} &= r_\text{p}\cdot\cos(\varphi^\prime), \\
  v_{\text{p}} &= r_\text{p}\cdot\sin(\varphi^\prime).\label{eq:perspective:end}
\end{align}

The generalized perspective projection function $\boldsymbol{\xi}_\text{p}: \mathcal{S} \rightarrow \mathbb{R}^2$ combines the described projection steps~\eqref{eq:perspective:begin}-\eqref{eq:perspective:end} in a joint expression taking the original spherical pixel coordinate $\vec{s}$ as input and returning the corresponding pixel coordinate $\vec{p}_\text{p}$ on the perspective image plane.

The inverse projection function $\boldsymbol{\xi}_\text{p}^{-1}: \mathbb{R}^2 \rightarrow \mathcal{S}$ is responsible for inverting the steps ~\eqref{eq:perspective:begin}-\eqref{eq:perspective:end} yielding the pixel coordinate $\vec{s}$ on the unit sphere based on the pixel coordinate $\vec{p}_\text{p}$ on the perspective image plane.

To obtain an exact reprojection from the perspective domain to the spherical domain, the binary variable $b_\text{vip}$ from~\eqref{eq:perspective:bvip} needs to be available as well.
In an implementation, $b_\text{vip}$ could be stored as side information or could be dynamically inferred from the original pixel position.
In further considerations, we drop it for reasons of clarity.

The inverse projection then entails the following steps.
First, the polar coordinates $(r_\text{p}, \varphi^\prime)$ need to be calculated from the pixel coordinate $\vec{p}_\text{p}$
\begin{align}
  r_\text{p} &= \sqrt{u_\text{p}^2 + v_\text{p}^2}, \\
  \varphi^{\prime} &= \arctantwo(v_\text{p}, u_\text{p}).
\end{align}
The corresponding incident angle is then obtained using
\begin{align}
  \theta^{\prime} = \begin{cases}
                      \arctan \left( r_\text{p}/f \right) & \text{if $b_\text{vip} = 0$,} \\
                      \pi - \arctan \left( r_\text{p}/f \right) & \text{if $b_\text{vip} = 1$.}
  \end{cases}
\end{align}
Finally, the pixel coordinate $\vec{s}$ in the spherical domain results in
\begin{align}
  x &= -\cos(\theta^\prime), \\
  y &= \sin(\theta^\prime) \cos(\varphi^\prime), \\
  z &= -\sin(\theta^\prime) \sin(\varphi^\prime).
\end{align}

\subsection{Motion Model}\label{subsec:motion-model}

Based on the generalized perspective projection function~$\boldsymbol{\xi}_\text{p}$ and a given 360-degree projection function $\boldsymbol{\xi}_\text{o}$ such as the equirectangular projection function $\boldsymbol{\xi}_\text{erp}$ described in Section~\ref{subsec:projections}, the motion-plane-adaptive motion model is derived as follows.

In a first step, the original 360-degree pixel coordinate $\vec{p}_\text{o}$ (in, e.g., the ERP domain) is projected to the pixel coordinate $\vec{p}_\text{p}$ on the desired motion plane described by $\mat{R}$ using the reprojection function $\boldsymbol{\zeta}_{\mat{R}}: \mathbb{R}^2 \rightarrow \mathbb{R}^2$ defined as
\begin{align}
  \vec{p}_\text{p} = \boldsymbol{\zeta}_{\mat{R}}(\vec{p}_\text{o}) = \boldsymbol{\xi}_\text{p}\left( \mat{R}\boldsymbol{\xi}_\text{o}^{-1}(\vec{p}_\text{o}) \right),\label{eq:mpa-motion-model:step1}
\end{align}
where $\boldsymbol{\xi}_\text{o}^{-1}$ projects the original 360-degree pixel coordinate $\vec{p}_\text{o}$ to the unit sphere, the motion plane rotation matrix $\mat{R} \in \mathbb{R}^{3 \times 3}$ rotates the pixel coordinate on the unit sphere according to the desired motion plane orientation, and $\boldsymbol{\xi}_\text{p}$ then projects the pixel coordinate onto the motion plane.

In a second step, the translational motion according to the motion vector $\vec{t}$ is performed on the obtained motion plane yielding the moved pixel coordinate on the motion plane
\begin{align}
  \vec{p}_\text{p,m} = \vec{p}_\text{p} + \vec{t}.\label{eq:mpa-motion-model:step2}
\end{align}

In the final third step, the moved pixel coordinate $\vec{p}_\text{p,m}$ on the motion plane is projected back to the original 360-degree format to obtain the moved pixel coordinate in the 360-degree projection $\vec{p}_\text{o,m}$ using the inverse reprojection function $\boldsymbol{\zeta}_{\mat{R}}^{-1}$ yielding
\begin{align}
  \vec{p}_\text{o,m} = \boldsymbol{\zeta}^{-1}_{\mat{R}}(\vec{p}_\text{p,m}) = \boldsymbol{\xi}_\text{o}\left( \mat{R}^{-1}\boldsymbol{\xi}_\text{p}^{-1}(\vec{p}_\text{p,m}) \right),\label{eq:mpa-motion-model:step3}
\end{align}
with $\mat{R}^{-1} = \mat{R}^T$.

Putting steps~\eqref{eq:mpa-motion-model:step1}-\eqref{eq:mpa-motion-model:step3} together, the overall motion model $\vec{m}_\text{mpa}$ is defined as
\begin{align}
  \vec{m}_\text{mpa}(\vec{p}_\text{o}, \vec{t}, \mat{R}) &= \boldsymbol{\zeta}^{-1}_{\mat{R}}\big(\boldsymbol{\zeta}_{\mat{R}}(\vec{p}_\text{o}) + \vec{t}\big) \label{eq:mpa-motion-model} \\
  &= \boldsymbol{\xi}_\text{o}\left( \mat{R}^{-1}\boldsymbol{\xi}_\text{p}^{-1}\left( \boldsymbol{\xi}_\text{p}\left( \mat{R}\boldsymbol{\xi}_\text{o}^{-1}(\vec{p}_\text{o}) \right) + \vec{t} \right)  \right).\nonumber
\end{align}
A schematic representation of the described motion model is shown in Fig.~\ref{fig:mpa-schematic}.
The figure also visualizes block motion for an exemplary block in an ERP-projected 360-degree image, where it is clearly visible that the proposed motion model is able to accurately replicate the distortions of the block in the ERP domain resulting from a translational motion on the street surface.

Rotations by multiples of 90 degrees around one or more axes can be formulated through simple transpositions of the 3D space coordinates, such that a suitably defined set of motion planes allows to considerably speed up the calculations for the inherent coordinate rotations.
We thus formulate a limited set of three rotation matrices leading to the motion planes
\begin{itemize}
  \setlength{\itemsep}{2pt}
  \item front/back: No rotation,
  \item left/right: $\pi/2$ around $z$-axis,
  \item top/bottom: $\pi/2$ around $y$-axis.
\end{itemize}
A potential encoder can then select the best matching motion plane through rate-distortion optimization.
Please note, however, that in general, the motion-plane-adaptive motion model is not limited to these motion planes and could employ arbitrary motion plane rotation matrices $\mat{R}$.

\subsection{Adapted Motion Vector Prediction}\label{subsec:mpamvp}

In the context of video coding, the motion information from spatially or temporally neighboring blocks is commonly reused in subsequently coded blocks to improve the overall compression efficiency.
As an example, video codecs often predict the motion vector for the current block from already available motion information of previously coded blocks by forming motion vector predictors (MVPs) at distinct candidate positions in spatially or temporally neighboring blocks~\cite{Wiegand2003, Sullivan2012, Bross2021a}.
The selected MVP is further refined at the encoder and only the index of the selected MVP as well as the motion vector difference (MVD) between the final estimated motion vector and the MVP needs to be signaled.
This greatly improves the compression efficiency of modern video codecs.

However, in the context of MPA, the motion information at the different candidate positions could be represented on motion planes differing from the investigated motion plane for the currently regarded block.
Hence, a method to efficiently translate motion information, i.e., motion vectors, between different motion planes is required.

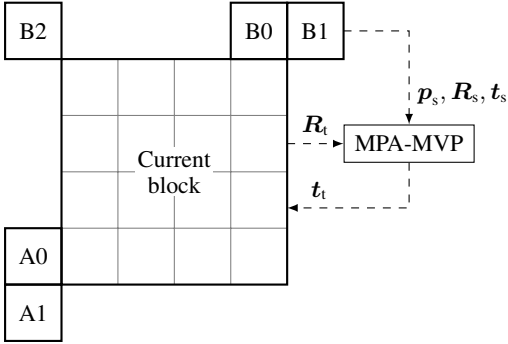
\begin{figure}[t]
  \centering
  \tikzstyle{arrow}=[-latex, dashed]
\begin{tikzpicture}[scale=3]
    \small
    \pgfmathsetmacro{\blocksize}{1}
    \pgfmathsetmacro{\pixelsize}{0.25}

    \coordinate (block_bl) at (0, 0);
    \path (block_bl) +(\blocksize, 0) coordinate (block_br);
    \path (block_bl) +(\blocksize, \blocksize) coordinate (block_tr);
    \path (block_bl) +(0, \blocksize) coordinate (block_tl);
    \path (block_bl) -- (block_tr) coordinate[midway] (block_center);
    \draw[step=0.25, gray, thin] (block_bl) grid (block_tr);
    \draw[thick] (block_bl) -- (block_br) -- (block_tr) -- (block_tl) -- cycle;
    \node[align=center, fill=white, inner sep=1pt] at (block_center) {Current\\block};

    \path (block_tr) +(-\pixelsize, 0) coordinate (b0_bl);
    \path (b0_bl) +(\pixelsize, 0) coordinate (b0_br);
    \path (b0_bl) +(\pixelsize, \pixelsize) coordinate (b0_tr);
    \path (b0_bl) +(0, \pixelsize) coordinate (b0_tl);
    \path (b0_bl) -- (b0_tr) coordinate[midway] (b0_center);
    \draw[thick] (b0_bl) -- (b0_br) -- (b0_tr) -- (b0_tl) -- cycle;
    \node at (b0_center) {B0};

    \path (block_tr) +(0, 0) coordinate (b1_bl);
    \path (b1_bl) +(\pixelsize, 0) coordinate (b1_br);
    \path (b1_bl) +(\pixelsize, \pixelsize) coordinate (b1_tr);
    \path (b1_bl) +(0, \pixelsize) coordinate (b1_tl);
    \path (b1_bl) -- (b1_tr) coordinate[midway] (b1_center);
    \draw[thick] (b1_bl) -- (b1_br) -- (b1_tr) -- (b1_tl) -- cycle;
    \node at (b1_center) {B1};

    \path (block_tl) +(-\pixelsize, 0) coordinate (b2_bl);
    \path (b2_bl) +(\pixelsize, 0) coordinate (b2_br);
    \path (b2_bl) +(\pixelsize, \pixelsize) coordinate (b2_tr);
    \path (b2_bl) +(0, \pixelsize) coordinate (b2_tl);
    \path (b2_bl) -- (b2_tr) coordinate[midway] (b2_center);
    \draw[thick] (b2_bl) -- (b2_br) -- (b2_tr) -- (b2_tl) -- cycle;
    \node at (b2_center) {B2};

    \path (block_bl) +(-\pixelsize, 0) coordinate (a0_bl);
    \path (a0_bl) +(\pixelsize, 0) coordinate (a0_br);
    \path (a0_bl) +(\pixelsize, \pixelsize) coordinate (a0_tr);
    \path (a0_bl) +(0, \pixelsize) coordinate (a0_tl);
    \path (a0_bl) -- (a0_tr) coordinate[midway] (a0_center);
    \draw[thick] (a0_bl) -- (a0_br) -- (a0_tr) -- (a0_tl) -- cycle;
    \node at (a0_center) {A0};

    \path (block_bl) +(-\pixelsize, -\pixelsize) coordinate (a1_bl);
    \path (a1_bl) +(\pixelsize, 0) coordinate (a1_br);
    \path (a1_bl) +(\pixelsize, \pixelsize) coordinate (a1_tr);
    \path (a1_bl) +(0, \pixelsize) coordinate (a1_tl);
    \path (a1_bl) -- (a1_tr) coordinate[midway] (a1_center);
    \draw[thick] (a1_bl) -- (a1_br) -- (a1_tr) -- (a1_tl) -- cycle;
    \node at (a1_center) {A1};

    \path (block_tr) +(\pixelsize, {-1.5*\pixelsize}) coordinate[anchor=west] node[anchor=west, draw=black, inner sep=4pt] (mpa_block) {MPA-MVP};
    \path (b1_tr) -- (b1_br) coordinate[midway] (b1_r);
    \draw[arrow] (b1_r) -| (mpa_block.north) node[midway, anchor=west, align=center, yshift=-24pt] {$\vec{p}_\text{s}, \mat{R}_\text{s}, \vec{t}_\text{s}$};
    \draw[arrow] (mpa_block.west -| block_tr) -- (mpa_block.west) node[midway, anchor=south, align=center] {$\mat{R}_\text{t}$};
    \path (mpa_block.south) +(0, {-0.8*\pixelsize}) coordinate (mpa_block_south_corner);
    \draw[arrow] (mpa_block.south) |- (mpa_block_south_corner -| block_tr) node[midway, anchor=south, align=center, xshift=-34pt] {$\vec{t}_\text{t}$};
\end{tikzpicture}
  \vspace*{-1em}
  \caption{Spatial motion vector predictor candidates in the H.266/VVC video coding standard~\cite{VVC-Draft10} with a schmematic illustration of MPA-MVP for candidate position B1 at pixel coordinate $\vec{p}_\text{s}$. With $\mat{R}_\text{s}$ describing the motion plane of the candidate position and $\mat{R}_\text{t}$ describing the motion plane of the current block, MPA-MVP translates the motion vector $\vec{t}_\text{s}$ from B1 to the corresponding motion vector predictor $\vec{t}_\text{t}$ for the current block.}
  \label{fig:mpa-mvp}
\end{figure}

According to Fig.~\ref{fig:mpa-mvp}, let's assume a motion vector $\vec{t}_\text{s}$ and a motion plane rotation matrix $\mat{R}_\text{s}$ are obtained at a source candidate pixel coordinate $\vec{p}_\text{s}$.
To transform the motion vector to a different target motion plane described by the rotation matrix $\mat{R}_\text{t}$, we need to find a motion vector $\vec{t}_\text{t}$ on the target motion plane that results in an identical pixel shift in the 360-degree domain as the motion vector $\vec{t}_\text{s}$ on the source motion plane, i.e.,
\begin{align}
  \vec{m}_\text{mpa}(\vec{p}_\text{s}, \vec{t}_\text{s}, \mat{R}_\text{s}) &\stackrel{!}{=} \vec{m}_\text{mpa}(\vec{p}_\text{s}, \vec{t}_\text{t}, \mat{R}_\text{t}). \label{eq:mpa-mvp-base}
\end{align}
The source pixel coordinate $\vec{p}_\text{s}$ is hereby used as an anchor for the motion plane translation.
By inserting~\eqref{eq:mpa-motion-model} into~\eqref{eq:mpa-mvp-base}, we can solve for the required motion vector on the target motion plane as
\begin{align}
  \vec{t}_{\text{t}} &= \boldsymbol{\zeta}_{\mat{R}_\text{t}}\left( \boldsymbol{\zeta}^{-1}_{\mat{R}_\text{s}}\left( \boldsymbol{\zeta}_{\mat{R}_\text{s}}(\vec{p}_\text{s}) + \vec{t}_\text{s} \right) \right) - \boldsymbol{\zeta}_{\mat{R}_\text{t}}(\vec{p}_\text{s}). \label{eq:mpa_to_mpa}
%
\end{align}
Thereby, the first term explains, where the candidate pixel coordinate moved by $\vec{t}_\text{s}$ on the source motion plane lies on the target motion plane, and the second term explains, where the \textit{"unmoved"} candidate pixel coordinate lies on the target motion plane.
Fig.~\ref{fig:mpa-mvp} visualizes the general procedure of the described motion-plane-adaptive motion vector prediction (MPA-MVP) for an exemplary candidate position B1.

Furthermore, to integrate MPA as an additional tool alongside the existing translational inter prediction procedure, MPA-MVP does not only need to be able to translate motion vectors between the different motion planes of MPA, but also between MPA and the classical translational motion model.
An equivalent motion vector can again be derived by setting up an equation similar to~\eqref{eq:mpa-mvp-base} for the translational and the motion-plane-adaptive motion model, and solving for the unknown motion vector by inserting~\eqref{eq:translational-motion-model} and~\eqref{eq:mpa-motion-model}.

In case the motion-plane-adaptive motion model is the source model, an equivalent motion vector $\vec{t}_\text{t}$ in the 360-degree domain for the classical translational motion model can then be derived based on the source motion vector $\vec{t}_\text{s}$ and motion plane described by $\mat{R}_\text{s}$ as
\begin{align}
  \vec{t}_{\text{t}} = \boldsymbol{\zeta}^{-1}_{\mat{R}_\text{s}}\left( \boldsymbol{\zeta}_{\mat{R}_\text{s}}(\vec{p}_\text{s}) + \vec{t}_\text{s} \right) - \vec{p}_{\text{s}} \label{eq:mpa_to_classic}
\end{align}

On the other hand, in case the classical translational motion model is the source model, an equivalent motion vector $\vec{t}_\text{t}$ on the motion plane described by $\mat{R}_\text{t}$ for the motion-plane-adaptive motion model can be derived based on the source motion vector $\vec{t}_\text{t}$ in the 360-degree domain as
\begin{align}
  \vec{t}_{\text{t}} &= \boldsymbol{\zeta}_{\mat{R}_\text{t}}(\vec{p}_\text{s} + \vec{t}_\text{s}) - \boldsymbol{\zeta}_{\mat{R}_\text{t}}(\vec{p}_\text{s}). \label{eq:classic_to_mpa}
\end{align}

The derived motion vectors can then serve as MVPs for the current block even if the motion model or motion plane of the candidate block and the current block do not match.

\subsection{Coding}\label{subsec:coding}

\begin{figure*}[t]
  \centering
  \scalebox{0.96}{\parbox{\linewidth}{\tikzstyle{block}=[draw, thick, minimum width=68pt, minimum height=30pt, align=center, node distance=42pt]
\tikzstyle{decision}=[draw, thick, diamond, minimum width=36pt, minimum height=36pt, align=center, node distance=34pt]
\tikzstyle{modified}=[anchor=center, draw, fill=white, rounded corners=2pt]
\tikzstyle{arrow}=[-latex]

\begin{tikzpicture}
	\footnotesize

	\node[block, fill=white!30!carolinablue] (signaling) at (0, 0) {Signaling};
	\node[block, fill=white!70!green, right = of signaling] (motion_derivation) {Motion Derivation};
	\node[block, right = of motion_derivation] (motion_modeling) {Motion Modeling};
	\node[block, right = of motion_modeling, align=center] (interpolation) {Reference Frame\\Sampling};

	\node[left = 17pt of signaling] (bitstream) {Bitstream};
	\node[right = 17pt of interpolation] (prediction) {Prediction};

	\path[fill=white!50!red, rounded corners=4pt] ([xshift=-8pt, yshift=18pt] motion_modeling.north west) rectangle ([xshift=8pt, yshift=-8pt] interpolation.south east) node[pos=.5, yshift=19pt] (motion_compensation) {\underline{Motion Compensation}};

	\node[block, fill=white!46!orange] at (motion_modeling) {Motion Modeling};
	\node[block, right = of motion_modeling, align=center, fill=white!50!better_yellow] (interpolation) {Reference Frame\\Sampling};

	\draw[arrow] (bitstream) -- (signaling);
	\draw[arrow] (signaling) -- (motion_derivation) node[midway, align=center] {Incomplete\\[2pt]MI};
	\draw[arrow] (motion_derivation) -- (motion_modeling) node[midway, align=center, xshift=-4pt] {Complete\\[2pt]MI};
	\draw[arrow] (motion_modeling) -- (interpolation);
	\draw[arrow] (interpolation) -- (prediction);

	\path (motion_modeling) -- (interpolation) node[midway] (motion_field_t) {};
	\node[below = 28pt of motion_field_t] (motion_field_b) {Motion Field};
	\draw[arrow] (motion_field_t.center) -- (motion_field_b);

	\coordinate (md_O) at (-1pt, -130pt);
	\node[decision, fill=white!90!green] (merge_check) at (md_O) {Merge?};
	\node[block, fill=white!90!green, above right = 25pt and 34pt of md_O] (merge_1) {Merge Candidate\\Adoption};
	\node[block, fill=white!90!green, below right = 25pt and 34pt of md_O] (merge_0) {MVP + MVD};
	\coordinate[below right = 25pt and 17pt of merge_1] (md_circle) {};

	\path[fill=white!70!green, rounded corners=4pt] ([xshift=-64pt, yshift=22pt] merge_1.north west) rectangle ([xshift=10pt, yshift=-66pt] md_circle.south east) node[pos=.5, yshift=62pt] (motion_derivation_encapsulating) {\underline{Motion Derivation}};

	\node[decision, fill=white!90!green] (merge_check) at (md_O) {\scriptsize Merge?};
	\node[block, fill=white!90!green, above right = 25pt and 34pt of md_O] (merge_1) {Merge Candidate\\Adoption};
	\node[block, fill=white!90!green, below right = 25pt and 34pt of md_O] (merge_0) {MVP + MVD};
	\node[left = 25pt of merge_check, align=center] (incomplete_mi) {Incomplete\\MI};
	\fill[black] (md_circle) circle (1pt);
	\node[right = 17pt of md_circle, align=center] (complete_mi) {Complete\\MI};
	\draw[arrow] (merge_check) |- (merge_1) node[midway, anchor=south] {\scriptsize Merge=1};
	\draw[arrow] (merge_check) |- (merge_0) node[midway, anchor=north] {\scriptsize Merge=0};
	\draw[arrow] (incomplete_mi) -- (merge_check);
	\draw (merge_1) -| (md_circle);
	\draw (merge_0) -| (md_circle);
	\draw[arrow] (md_circle) -- (complete_mi);

	\coordinate[right = 36pt of complete_mi.east, anchor=west] (mm_O);
	\node[decision, fill=white!75!orange] (mpa_check) at (mm_O) {\scriptsize MPA?};
	\node[decision, fill=white!75!orange, above right = 11pt and 30pt of mm_O, align=center] (affine_check) {\scriptsize Affine?};
	\node[block, fill=white!75!orange, above right = 25pt and 64pt of mm_O] (translational) {Translational};
	\node[block, fill=white!75!orange, right = 64pt of mm_O] (affine) {Affine};
	\node[block, fill=white!75!orange, below right = 25pt and 64pt of mm_O, align=center] (mpa) {Motion-Plane-\\Adaptive};
	\coordinate[below right = 25pt and 17pt of translational] (mm_circle) {};

	\path[fill=white!46!orange, rounded corners=4pt] ([xshift=-90pt, yshift=22pt] translational.north west) rectangle ([xshift=10pt, yshift=-66pt] mm_circle.south east) node[pos=.5, yshift=62pt] (motion_modeling_encapsulating) {\underline{Motion Modeling}};

  \node[decision, fill=white!75!orange] (mpa_check) at (mm_O) {\scriptsize MPA?};
	\node[decision, fill=white!75!orange, above right = 11pt and 30pt of mm_O, align=center] (affine_check) {\scriptsize Affine?};
	\node[block, fill=white!75!orange, above right = 25pt and 64pt of mm_O] (translational) {Translational};
	\node[block, fill=white!75!orange, right = 64pt of mm_O] (affine) {Affine};
	\node[block, fill=white!75!orange, below right = 25pt and 64pt of mm_O, align=center] (mpa) {Motion-Plane-\\Adaptive};
	\coordinate[below right = 25pt and 17pt of translational] (mm_circle) {};

	\fill[black] (mm_circle) circle (1pt);
	\node[right = 17pt of mm_circle, align=center] (mm_motion_field) {Motion\\Field};
	\draw[arrow] (mpa_check) |- (mpa) node[midway, anchor=north] {\scriptsize MPA=1};
	\draw[arrow] (mpa_check) |- (affine_check) node[midway, anchor=south] {\scriptsize MPA=0};
	\draw[arrow] (affine_check) |- (affine) node[midway, anchor=north] {\scriptsize Affine=1};
	\draw[arrow] (affine_check) |- (translational) node[midway, anchor=south] {\scriptsize Affine=0};
	\draw[arrow] (complete_mi) -- (mpa_check);
	\draw (translational) -| (mm_circle);
	\draw (affine) -- (mm_circle);
	\draw (mpa) -| (mm_circle);
	\draw[arrow] (mm_circle) -- (mm_motion_field);

	\node[modified, fill=white!30!carolinablue] at (signaling.south) {\tiny Extended};
	\node[modified, fill=white!70!green] at (motion_derivation.south) {\tiny Extended};
	\node[modified, fill=white!46!orange] at (motion_modeling.south) {\tiny Extended};
	\node[modified, fill=white!90!green] at (merge_1.south) {\tiny Extended};
	\node[modified, fill=white!90!green] at (merge_0.south) {\tiny Extended};
	\node[modified, fill=white!75!orange] at (mpa_check.south) {\tiny New};
	\node[modified, fill=white!75!orange] at (mpa.south) {\tiny New};

	\draw[dashed, thin] (motion_derivation.south west) -- ([xshift=-80pt, yshift=2pt] motion_derivation_encapsulating.north);
	\draw[dashed, thin] (motion_derivation.south east) -- ([xshift=80pt, yshift=0pt] motion_derivation_encapsulating.north);
	\draw[dashed, thin] (motion_modeling.south west) -- ([xshift=-92pt, yshift=0pt] motion_modeling_encapsulating.north);
	\draw[dashed, thin] (motion_modeling.south east) -- ([xshift=92pt, yshift=2pt] motion_modeling_encapsulating.north);
\end{tikzpicture}}}
  \vspace*{-1em}
  \caption{Schematic representation of the decoder-side inter prediction pipeline in the H.266/VVC video coding standard including the proposed MPA tool. Components that need to be extended or added with respect to the original H.266/VVC inter prediction pipeline are labeled explicitly. For details on the performed extensions and additions, please see the text.}
  \label{fig:inter-prediction-pipeline}
\end{figure*}
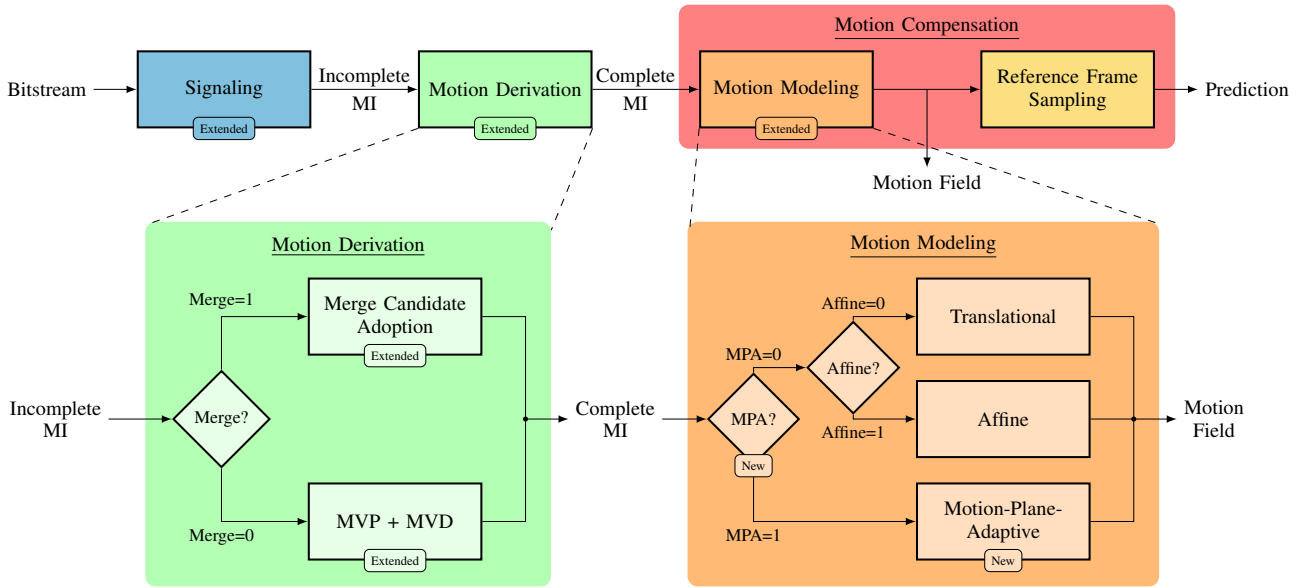

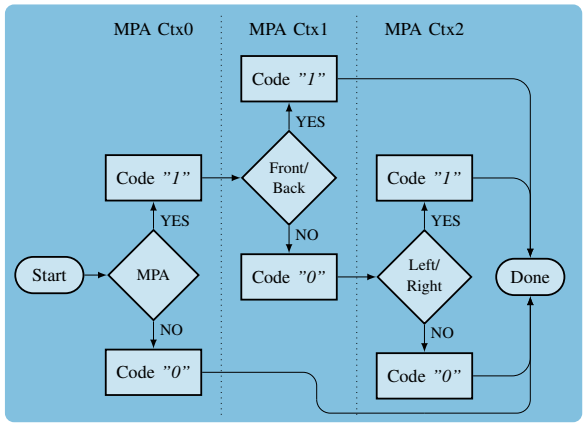
\begin{figure}[t]
  \centering
  \scalebox{0.86}{\parbox{\columnwidth}{\tikzstyle{block}=[draw, thick, minimum width=42pt, minimum height=20pt, align=center, node distance=42pt]
\tikzstyle{decision}=[draw, thick, diamond, minimum width=40pt, minimum height=40pt, align=center, node distance=34pt]
\tikzstyle{modified}=[anchor=center, draw, fill=white, rounded corners=2pt]
\tikzstyle{arrow}=[-latex]

\begin{tikzpicture}
	\footnotesize

	\node[decision, fill=white!100!carolinablue] (mpa) at (0, 0) {\scriptsize MPA};
	\node[block, above = 12pt of mpa, fill=white!100!carolinablue] (mpa_1) {Code \textit{"1"}};
	\node[block, below = 12pt of mpa, fill=white!100!carolinablue] (mpa_0) {Code \textit{"0"}};

	\node[decision, right = 17pt of mpa_1, fill=white!100!carolinablue] (front_back) {\scriptsize Front/\\\scriptsize Back};
	\node[block, above=12pt of front_back, fill=white!100!carolinablue] (front_back_1) {Code \textit{"1"}};
	\node[block, below=12pt of front_back, fill=white!100!carolinablue] (front_back_0) {Code \textit{"0"}};

	\node[decision, right = 17pt of front_back_0, fill=white!100!carolinablue] (left_right) {\scriptsize Left/\\\scriptsize Right};
	\node[block, above=12pt of left_right, fill=white!100!carolinablue] (left_right_1) {Code \textit{"1"}};
	\node[block, below=12pt of left_right, fill=white!100!carolinablue] (left_right_0) {Code \textit{"0"}};

	\node[draw, thick, rounded corners=8pt, minimum width=30pt, minimum height=16pt, left = 10pt of mpa, fill=white!100!carolinablue] (start) {Start};
	\node[draw, thick, rounded corners=8pt, minimum width=30pt, minimum height=16pt, right = 10pt of left_right, fill=white!100!carolinablue] (done) {Done};

	\draw[arrow] (start) -- (mpa);

	\draw[arrow] (mpa) -- (mpa_1) node[pos=0.3, anchor=west] {\scriptsize YES};
	\draw[arrow] (mpa) -- (mpa_0) node[pos=0.3, anchor=west] {\scriptsize NO};
	\draw[arrow] (mpa_1) -- (front_back);
	\draw[arrow] (front_back) -- (front_back_1) node[pos=0.3, anchor=west] {\scriptsize YES};
	\draw[arrow] (front_back) -- (front_back_0) node[pos=0.3, anchor=west] {\scriptsize NO};
	\draw[arrow] (front_back_0) -- (left_right);
	\draw[arrow] (left_right) -- (left_right_1) node[pos=0.3, anchor=west] {\scriptsize YES};
	\draw[arrow] (left_right) -- (left_right_0) node[pos=0.3, anchor=west] {\scriptsize NO};

	\draw[arrow, rounded corners=8pt] (front_back_1) -| (done);
	\draw[arrow, rounded corners=8pt] (left_right_1) -| (done);
	\draw[arrow, rounded corners=8pt] (left_right_0) -| (done);

	\coordinate[left = 26pt of left_right_0.south west] (mpa_0_p1);
	\coordinate[below = 6pt of left_right_0.south] (mpa_0_p2);
	\draw[rounded corners = 6pt] (mpa_0) -| (mpa_0_p1);
	\draw[rounded corners = 6pt] (mpa_0_p1) |- (mpa_0_p2);
	\draw[arrow, rounded corners = 6pt] (mpa_0_p2) -| (done);

	\node[above = 6pt of front_back_1] (mpa_ctx_1) {MPA Ctx1};
	\draw (mpa_ctx_1 -| mpa) node (mpa_ctx_0) {MPA Ctx0};
	\draw (mpa_ctx_1 -| left_right) node (mpa_ctx_2) {MPA Ctx2};
	\path (mpa_ctx_0.north) -- (mpa_ctx_1.north) coordinate[midway] (line_ctx_01);
	\path (mpa_ctx_1.north) -- (mpa_ctx_2.north) coordinate[midway] (line_ctx_12);
	\coordinate[below = 2pt of mpa_0_p2] (lower);
	\draw[dotted, thin] (line_ctx_01) -- (line_ctx_01 |- lower);
	\draw[dotted, thin] (line_ctx_12) -- (line_ctx_12 |- lower);

	\path[fill=white!30!carolinablue, rounded corners=4pt] ([xshift=-45pt, yshift=5pt] mpa_ctx_0.north west) rectangle ([xshift=48pt, yshift=-10pt] left_right_0.south east) node[pos=.5, yshift=0pt] (signaling_syntax_encapsulating) {};

	\node[decision, fill=white!70!carolinablue] (mpa) at (0, 0) {\scriptsize MPA};
	\node[block, above = 12pt of mpa, fill=white!70!carolinablue] (mpa_1) {Code \textit{"1"}};
	\node[block, below = 12pt of mpa, fill=white!70!carolinablue] (mpa_0) {Code \textit{"0"}};

	\node[decision, right = 17pt of mpa_1, fill=white!70!carolinablue] (front_back) {\scriptsize Front/\\\scriptsize Back};
	\node[block, above=12pt of front_back, fill=white!70!carolinablue] (front_back_1) {Code \textit{"1"}};
	\node[block, below=12pt of front_back, fill=white!70!carolinablue] (front_back_0) {Code \textit{"0"}};

	\node[decision, right = 17pt of front_back_0, fill=white!70!carolinablue] (left_right) {\scriptsize Left/\\\scriptsize Right};
	\node[block, above=12pt of left_right, fill=white!70!carolinablue] (left_right_1) {Code \textit{"1"}};
	\node[block, below=12pt of left_right, fill=white!70!carolinablue] (left_right_0) {Code \textit{"0"}};

	\node[draw, thick, rounded corners=8pt, minimum width=30pt, minimum height=16pt, left = 10pt of mpa, fill=white!70!carolinablue] (start) {Start};
	\node[draw, thick, rounded corners=8pt, minimum width=30pt, minimum height=16pt, right = 10pt of left_right, fill=white!70!carolinablue] (done) {Done};

	\draw[arrow] (start) -- (mpa);

	\draw[arrow] (mpa) -- (mpa_1) node[pos=0.3, anchor=west] {\scriptsize YES};
	\draw[arrow] (mpa) -- (mpa_0) node[pos=0.3, anchor=west] {\scriptsize NO};
	\draw[arrow] (mpa_1) -- (front_back);
	\draw[arrow] (front_back) -- (front_back_1) node[pos=0.3, anchor=west] {\scriptsize YES};
	\draw[arrow] (front_back) -- (front_back_0) node[pos=0.3, anchor=west] {\scriptsize NO};
	\draw[arrow] (front_back_0) -- (left_right);
	\draw[arrow] (left_right) -- (left_right_1) node[pos=0.3, anchor=west] {\scriptsize YES};
	\draw[arrow] (left_right) -- (left_right_0) node[pos=0.3, anchor=west] {\scriptsize NO};

	\draw[arrow, rounded corners=8pt] (front_back_1) -| (done);
	\draw[arrow, rounded corners=8pt] (left_right_1) -| (done);
	\draw[arrow, rounded corners=8pt] (left_right_0) -| (done);

	\coordinate[left = 26pt of left_right_0.south west] (mpa_0_p1);
	\coordinate[below = 6pt of left_right_0.south] (mpa_0_p2);
	\draw[rounded corners = 6pt] (mpa_0) -| (mpa_0_p1);
	\draw[rounded corners = 6pt] (mpa_0_p1) |- (mpa_0_p2);
	\draw[arrow, rounded corners = 6pt] (mpa_0_p2) -| (done);

	\node[above = 6pt of front_back_1] (mpa_ctx_1) {MPA Ctx1};
	\draw (mpa_ctx_1 -| mpa) node (mpa_ctx_0) {MPA Ctx0};
	\draw (mpa_ctx_1 -| left_right) node (mpa_ctx_2) {MPA Ctx2};
	\path (mpa_ctx_0.north) -- (mpa_ctx_1.north) coordinate[midway] (line_ctx_01);
	\path (mpa_ctx_1.north) -- (mpa_ctx_2.north) coordinate[midway] (line_ctx_12);
	\coordinate[below = 2pt of mpa_0_p2] (lower);
	\draw[dotted, thin] (line_ctx_01) -- (line_ctx_01 |- lower);
	\draw[dotted, thin] (line_ctx_12) -- (line_ctx_12 |- lower);

\end{tikzpicture}}}
  \vspace*{-1em}
  \caption{Signaling structure for the additional syntax elements required by MPA. Each column uses a dedicated context model for CABAC entropy coding denoted by \textit{MPA Ctx0} - \textit{Ctx2}.}
  \label{fig:mpa-signaling-syntax}
\end{figure}

To evaluate the application of MPA to an actual video codec, we integrate the proposed concept as an additional tool into the state-of-the-art H.266/VVC video coding standard.
The implementation is based on the VVC reference software \mbox{VTM-14.2}~\cite{VTM-14.2, Browne2021}.
To facilitate the adaption of dependent tools in the coding pipeline, our studies are currently focused on the luma component.

Fig.~\ref{fig:inter-prediction-pipeline} shows a schematic representation of the decoder-side inter prediction pipeline of H.266/VVC for a single inter predicted coding unit (CU) and highlights where extensions and additions are required in order to support MPA.
For reasons of clarity, not all inter prediction tools available in the H.266/VVC video coding standard are depicted.

In the following, we discuss the different inter prediction steps and explain our proposed extensions and additions to support MPA.
Any required encoder-side adaptions are introduced at the appropriate positions.

\subsubsection*{Signaling}
If a merge mode is signaled or the current CU is coded using an affine motion model, no adaptions to the signaling procedure are required for MPA.
If no merge mode is signaled and the current CU is not coded using an affine motion model, the integration of MPA requires additional information to be signaled.
In reference to Subclause 7.3.11.5 in the VVC standard draft~\cite{VVC-Draft10}, the MPA-specific information (MPA flag, motion plane index) is signaled after the \textbf{inter\_affine\_flag} provided the affine motion model is not active.
Fig.~\ref{fig:mpa-signaling-syntax} describes the signaling structure of the corresponding syntax elements required for MPA.
A first bin encodes an MPA flag to indicate whether MPA or the traditional translational motion model is used.
In case MPA is active, a second bin denotes whether the front/back motion plane is used, and, if this is not the case, a third bin denotes whether the left/right or the top/bottom motion plane is used.
Similar to the existing motion information (MI) in the H.266/VVC video coding standard, the described information is signaled using the context-based binary arithmetic coder (CABAC)~\cite{Marpe2003} with dedicated context models, which are represented by \textit{MPA Ctx0} - \textit{Ctx2} in Fig.~\ref{fig:mpa-signaling-syntax}.
In case of bi-prediction, i.e., two predictions with different MI are averaged to form an overall prediction for the current CU, both predictions share the same motion plane, such that the motion plane index needs to be signaled only once.

\subsubsection*{Motion Derivation}
After the signaling step shaded blue in Fig.~\ref{fig:inter-prediction-pipeline}, the MI is incomplete and needs to be completed in a motion derivation step.
As such, a CU can either apply a merge mode to take over all MI from a spatially or temporally neighboring CU, or explicitly signal the MI whereby only a motion vector difference (MVD) between the actual motion vector and the selected MVP is transmitted.
Hence, in both cases, not all MI required for motion compensation is available after signaling, as at least the motion vector needs to be derived regardless of whether a merge mode is applied or not.
The green box in the bottom row of Fig.~\ref{fig:inter-prediction-pipeline} describes these two cases where the specific algorithm for motion derivation depends on the value of the merge flag.

If a merge mode is applied (Merge=1),
support for MPA is added by extending the merge candidate information by a flag indicating the used inter prediction technique (classical or MPA) and, in case of MPA, an index indicating the selected motion plane.
The set of rules for the candidate list derivation remains largely unchanged.
The only exception are pairwise average merge candidates, where an additional restriction is formulated that requires both averaging candidates to share the same motion plane.

If the MI is explicitly signaled (Merge=0),
support for MPA is added by employing MPA-MVP according to~\eqref{eq:mpa_to_mpa},~\eqref{eq:mpa_to_classic}, and~\eqref{eq:classic_to_mpa} as described in Subsection~\ref{subsec:mpamvp}.
The adapted motion vector prediction scheme allows to efficiently switch between different motion planes and motion models without losing valuable prediction information from neighboring CUs.
By translating motion vectors from the motion-plane-adaptive motion model to the translational motion model using~\eqref{eq:mpa_to_classic}, MPA-coded CUs can also be used for the construction of affine MVPs and merge candidates for the affine motion model\cite{Li2018, Zhang2019}.

\subsubsection*{Motion Compensation}
After the motion derivation step, the motion compensation step generates a prediction for the current CU based on a known reference frame and the completed MI.
As shown in the red highlighted area in Fig.~\ref{fig:inter-prediction-pipeline}, this procedure can be further subdivided into a motion modeling step shaded orange and a reference frame sampling step shaded yellow.
The motion modeling step is responsible for generating a pixel-wise motion field describing the resulting pixel shifts after applying the respective motion models.

To add support for MPA, this step needs to be extended to include the novel motion-plane-adaptive motion model as visualized in the orange box in the bottom row of Fig.~\ref{fig:inter-prediction-pipeline}.
Without MPA, motion modeling has to decide between the translational and affine motion models based on a flag \textit{Affine} in the MI.
With the introduction of MPA, a further decision between the existing motion models and the proposed motion-plane-adaptive motion model is added based on a flag \textit{MPA} in the MI.
If MPA is selected for the current CU (MPA=1), the motion-plane-adaptive motion model is applied according to~\eqref{eq:mpa-motion-model}.

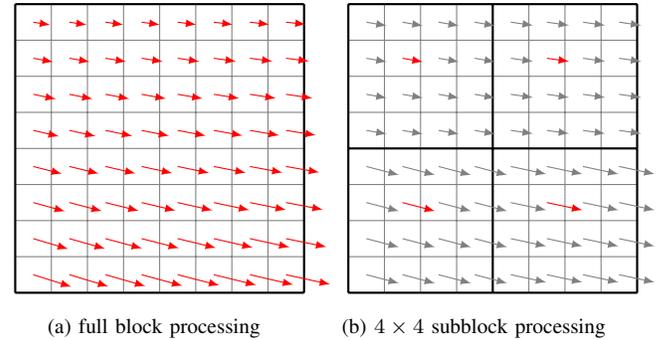
\begin{figure}[t]
  \centering
  \hfill
  \subfloat{%
    \parbox{0.45\columnwidth}{\centering\begin{tikzpicture}[scale=0.48]%
    \draw[step=1, gray, thin] (0, 0) grid (8, 8);
    \draw[step=8, black, thick] (0, 0) grid (8, 8);
    \draw[-latex, red](0.5, 7.5) -- +(0.47264421818617275, -0.07602724126275291);
    \draw[-latex, red](1.5, 7.5) -- +(0.48473742972069545, -0.07409388500676925);
    \draw[-latex, red](2.5, 7.5) -- +(0.4965059216908798, -0.07211435797496221);
    \draw[-latex, red](3.5, 7.5) -- +(0.5079423366989971, -0.07009012854369416);
    \draw[-latex, red](4.5, 7.5) -- +(0.5190395755618965, -0.06802269160363607);
    \draw[-latex, red](5.5, 7.5) -- +(0.5297908012862195, -0.0659135670338884);
    \draw[-latex, red](6.5, 7.5) -- +(0.5401894427482034, -0.06376429817046585);
    \draw[-latex, red](7.5, 7.5) -- +(0.5502291980786481, -0.0615764502721646);
    \draw[-latex, red](0.5, 6.5) -- +(0.5516996848699932, -0.10216015891281245);
    \draw[-latex, red](1.5, 6.5) -- +(0.5657665594726552, -0.09954177764325323);
    \draw[-latex, red](2.5, 6.5) -- +(0.5794508406337513, -0.0968615406157852);
    \draw[-latex, red](3.5, 6.5) -- +(0.5927440004359056, -0.09412148227727642);
    \draw[-latex, red](4.5, 6.5) -- +(0.605637824300311, -0.09132367186072088);
    \draw[-latex, red](5.5, 6.5) -- +(0.618124415402491, -0.08847021116052645);
    \draw[-latex, red](6.5, 6.5) -- +(0.6301961987081369, -0.08556323230556018);
    \draw[-latex, red](7.5, 6.5) -- +(0.6418459246312807, -0.08260489553383933);
    \draw[-latex, red](0.5, 5.5) -- +(0.6320275447772161, -0.13203430359712057);
    \draw[-latex, red](1.5, 5.5) -- +(0.6480852059526626, -0.1286235905526041);
    \draw[-latex, red](2.5, 5.5) -- +(0.663700434648976, -0.12513318449486235);
    \draw[-latex, red](3.5, 5.5) -- +(0.6788635354675989, -0.12156579453065602);
    \draw[-latex, red](4.5, 5.5) -- +(0.6935651859200561, -0.11792417351214794);
    \draw[-latex, red](5.5, 5.5) -- +(0.7077964411937728, -0.11421111493351645);
    \draw[-latex, red](6.5, 5.5) -- +(0.7215487384420441, -0.11042944983246343);
    \draw[-latex, red](7.5, 5.5) -- +(0.7348139006019607, -0.10658204370158728);
    \draw[-latex, red](0.5, 4.5) -- +(0.7137609705374245, -0.1655797866575483);
    \draw[-latex, red](1.5, 4.5) -- +(0.7318290118158671, -0.1612688604387216);
    \draw[-latex, red](2.5, 4.5) -- +(0.7493926702024453, -0.1568583216107811);
    \draw[-latex, red](3.5, 4.5) -- +(0.7664410867116487, -0.15235166802390268);
    \draw[-latex, red](4.5, 4.5) -- +(0.7829638395773202, -0.14775245071024584);
    \draw[-latex, red](5.5, 4.5) -- +(0.7989509492611716, -0.14306426970366884);
    \draw[-latex, red](6.5, 4.5) -- +(0.8143928828767741, -0.1382907698755247);
    \draw[-latex, red](7.5, 4.5) -- +(0.8292805580349977, -0.13343563679484305);
    \draw[-latex, red](0.5, 3.5) -- +(0.79704132500421, -0.20271353382638324);
    \draw[-latex, red](1.5, 3.5) -- +(0.8171418872870958, -0.19739431034829805);
    \draw[-latex, red](2.5, 3.5) -- +(0.8366738452226997, -0.19195355039820752);
    \draw[-latex, red](3.5, 3.5) -- +(0.8556251812376786, -0.18639565966663554);
    \draw[-latex, red](4.5, 3.5) -- +(0.8739843843460846, -0.18072510670900213);
    \draw[-latex, red](5.5, 3.5) -- +(0.8917404552736116, -0.17494641747189235);
    \draw[-latex, red](6.5, 3.5) -- +(0.9088829108746779, -0.16906416985356712);
    \draw[-latex, red](7.5, 3.5) -- +(0.9254017878522802, -0.16308298830935825);
    \draw[-latex, red](0.5, 2.5) -- +(0.8820192129131242, -0.24333908429864548);
    \draw[-latex, red](1.5, 2.5) -- +(0.9041770726918571, -0.23690366699933632);
    \draw[-latex, red](2.5, 2.5) -- +(0.9256996609721057, -0.2303228709128234);
    \draw[-latex, red](3.5, 2.5) -- +(0.946573807990751, -0.22360213313367863);
    \draw[-latex, red](4.5, 2.5) -- +(0.9667869253539606, -0.21674696330060783);
    \draw[-latex, red](5.5, 2.5) -- +(0.9863270111272697, -0.2097629365964991);
    \draw[-latex, red](6.5, 2.5) -- +(1.0051826540816649, -0.2026556868086601);
    \draw[-latex, red](7.5, 2.5) -- +(1.0233430371101917, -0.19543089946312478);
    \draw[-latex, red](0.5, 1.5) -- +(0.9688556551686645, -0.28734644794969594);
    \draw[-latex, red](1.5, 1.5) -- +(0.9930983240318312, -0.27968753389070283);
    \draw[-latex, red](2.5, 1.5) -- +(1.0166364171626354, -0.2718575733900252);
    \draw[-latex, red](3.5, 1.5) -- +(1.0394556207537733, -0.26386316236934704);
    \draw[-latex, red](4.5, 1.5) -- +(1.0615422829551449, -0.2557109787035188);
    \draw[-latex, red](5.5, 1.5) -- +(1.0828834187539542, -0.24740777344700574);
    \draw[-latex, red](6.5, 1.5) -- +(1.1034667138585177, -0.23896036215573133);
    \draw[-latex, red](7.5, 1.5) -- +(1.1232805276067437, -0.23037561632251025);
    \draw[-latex, red](0.5, 0.5) -- +(1.0577234058611082, -0.3346120249495755);
    \draw[-latex, red](1.5, 0.5) -- +(1.084081241729055, -0.32562332398470345);
    \draw[-latex, red](2.5, 0.5) -- +(1.109662350431293, -0.31643618352211433);
    \draw[-latex, red](3.5, 0.5) -- +(1.134451284422721, -0.30705848893110216);
    \draw[-latex, red](4.5, 0.5) -- +(1.1584333449429265, -0.29749821639304624);
    \draw[-latex, red](5.5, 0.5) -- +(1.1815945864803097, -0.28776342209356187);
    \draw[-latex, red](6.5, 0.5) -- +(1.2039218200715196, -0.2778622315569322);
    \draw[-latex, red](7.5, 0.5) -- +(1.2254026154653388, -0.2678028291453245);
\end{tikzpicture}%
}\vspace*{-2cm}}
  \hfill
  \subfloat{%
    \parbox{0.45\columnwidth}{\centering\begin{tikzpicture}[scale=0.48]%
    \draw[step=1, gray, thin] (0, 0) grid (8, 8);
    \draw[step=4, black, thick] (0, 0) grid (8, 8);
    \draw[-latex, red](1.5, 6.5) -- +(0.5657665594726552, -0.09954177764325323);
    \draw[-latex, gray, thin](0.5, 7.5) -- +(0.5657665594726552, -0.09954177764325323);
    \draw[-latex, gray, thin](0.5, 6.5) -- +(0.5657665594726552, -0.09954177764325323);
    \draw[-latex, gray, thin](0.5, 5.5) -- +(0.5657665594726552, -0.09954177764325323);
    \draw[-latex, gray, thin](0.5, 4.5) -- +(0.5657665594726552, -0.09954177764325323);
    \draw[-latex, gray, thin](1.5, 7.5) -- +(0.5657665594726552, -0.09954177764325323);
    \draw[-latex, gray, thin](1.5, 5.5) -- +(0.5657665594726552, -0.09954177764325323);
    \draw[-latex, gray, thin](1.5, 4.5) -- +(0.5657665594726552, -0.09954177764325323);
    \draw[-latex, gray, thin](2.5, 7.5) -- +(0.5657665594726552, -0.09954177764325323);
    \draw[-latex, gray, thin](2.5, 6.5) -- +(0.5657665594726552, -0.09954177764325323);
    \draw[-latex, gray, thin](2.5, 5.5) -- +(0.5657665594726552, -0.09954177764325323);
    \draw[-latex, gray, thin](2.5, 4.5) -- +(0.5657665594726552, -0.09954177764325323);
    \draw[-latex, gray, thin](3.5, 7.5) -- +(0.5657665594726552, -0.09954177764325323);
    \draw[-latex, gray, thin](3.5, 6.5) -- +(0.5657665594726552, -0.09954177764325323);
    \draw[-latex, gray, thin](3.5, 5.5) -- +(0.5657665594726552, -0.09954177764325323);
    \draw[-latex, gray, thin](3.5, 4.5) -- +(0.5657665594726552, -0.09954177764325323);
    \draw[-latex, red](5.5, 6.5) -- +(0.618124415402491, -0.08847021116052645);
    \draw[-latex, gray, thin](4.5, 7.5) -- +(0.618124415402491, -0.08847021116052645);
    \draw[-latex, gray, thin](4.5, 6.5) -- +(0.618124415402491, -0.08847021116052645);
    \draw[-latex, gray, thin](4.5, 5.5) -- +(0.618124415402491, -0.08847021116052645);
    \draw[-latex, gray, thin](4.5, 4.5) -- +(0.618124415402491, -0.08847021116052645);
    \draw[-latex, gray, thin](5.5, 7.5) -- +(0.618124415402491, -0.08847021116052645);
    \draw[-latex, gray, thin](5.5, 5.5) -- +(0.618124415402491, -0.08847021116052645);
    \draw[-latex, gray, thin](5.5, 4.5) -- +(0.618124415402491, -0.08847021116052645);
    \draw[-latex, gray, thin](6.5, 7.5) -- +(0.618124415402491, -0.08847021116052645);
    \draw[-latex, gray, thin](6.5, 6.5) -- +(0.618124415402491, -0.08847021116052645);
    \draw[-latex, gray, thin](6.5, 5.5) -- +(0.618124415402491, -0.08847021116052645);
    \draw[-latex, gray, thin](6.5, 4.5) -- +(0.618124415402491, -0.08847021116052645);
    \draw[-latex, gray, thin](7.5, 7.5) -- +(0.618124415402491, -0.08847021116052645);
    \draw[-latex, gray, thin](7.5, 6.5) -- +(0.618124415402491, -0.08847021116052645);
    \draw[-latex, gray, thin](7.5, 5.5) -- +(0.618124415402491, -0.08847021116052645);
    \draw[-latex, gray, thin](7.5, 4.5) -- +(0.618124415402491, -0.08847021116052645);
    \draw[-latex, red](1.5, 2.5) -- +(0.9041770726918571, -0.23690366699933632);
    \draw[-latex, gray, thin](0.5, 3.5) -- +(0.9041770726918571, -0.23690366699933632);
    \draw[-latex, gray, thin](0.5, 2.5) -- +(0.9041770726918571, -0.23690366699933632);
    \draw[-latex, gray, thin](0.5, 1.5) -- +(0.9041770726918571, -0.23690366699933632);
    \draw[-latex, gray, thin](0.5, 0.5) -- +(0.9041770726918571, -0.23690366699933632);
    \draw[-latex, gray, thin](1.5, 3.5) -- +(0.9041770726918571, -0.23690366699933632);
    \draw[-latex, gray, thin](1.5, 1.5) -- +(0.9041770726918571, -0.23690366699933632);
    \draw[-latex, gray, thin](1.5, 0.5) -- +(0.9041770726918571, -0.23690366699933632);
    \draw[-latex, gray, thin](2.5, 3.5) -- +(0.9041770726918571, -0.23690366699933632);
    \draw[-latex, gray, thin](2.5, 2.5) -- +(0.9041770726918571, -0.23690366699933632);
    \draw[-latex, gray, thin](2.5, 1.5) -- +(0.9041770726918571, -0.23690366699933632);
    \draw[-latex, gray, thin](2.5, 0.5) -- +(0.9041770726918571, -0.23690366699933632);
    \draw[-latex, gray, thin](3.5, 3.5) -- +(0.9041770726918571, -0.23690366699933632);
    \draw[-latex, gray, thin](3.5, 2.5) -- +(0.9041770726918571, -0.23690366699933632);
    \draw[-latex, gray, thin](3.5, 1.5) -- +(0.9041770726918571, -0.23690366699933632);
    \draw[-latex, gray, thin](3.5, 0.5) -- +(0.9041770726918571, -0.23690366699933632);
    \draw[-latex, red](5.5, 2.5) -- +(0.9863270111272697, -0.2097629365964991);
    \draw[-latex, gray, thin](4.5, 3.5) -- +(0.9863270111272697, -0.2097629365964991);
    \draw[-latex, gray, thin](4.5, 2.5) -- +(0.9863270111272697, -0.2097629365964991);
    \draw[-latex, gray, thin](4.5, 1.5) -- +(0.9863270111272697, -0.2097629365964991);
    \draw[-latex, gray, thin](4.5, 0.5) -- +(0.9863270111272697, -0.2097629365964991);
    \draw[-latex, gray, thin](5.5, 3.5) -- +(0.9863270111272697, -0.2097629365964991);
    \draw[-latex, gray, thin](5.5, 1.5) -- +(0.9863270111272697, -0.2097629365964991);
    \draw[-latex, gray, thin](5.5, 0.5) -- +(0.9863270111272697, -0.2097629365964991);
    \draw[-latex, gray, thin](6.5, 3.5) -- +(0.9863270111272697, -0.2097629365964991);
    \draw[-latex, gray, thin](6.5, 2.5) -- +(0.9863270111272697, -0.2097629365964991);
    \draw[-latex, gray, thin](6.5, 1.5) -- +(0.9863270111272697, -0.2097629365964991);
    \draw[-latex, gray, thin](6.5, 0.5) -- +(0.9863270111272697, -0.2097629365964991);
    \draw[-latex, gray, thin](7.5, 3.5) -- +(0.9863270111272697, -0.2097629365964991);
    \draw[-latex, gray, thin](7.5, 2.5) -- +(0.9863270111272697, -0.2097629365964991);
    \draw[-latex, gray, thin](7.5, 1.5) -- +(0.9863270111272697, -0.2097629365964991);
    \draw[-latex, gray, thin](7.5, 0.5) -- +(0.9863270111272697, -0.2097629365964991);
\end{tikzpicture}%
}\vspace*{-2cm}}
  \hfill
  \\\vspace*{-2em}\setcounter{subfigure}{0}
  \subfloat[full block processing\label{fig:mpa-4x4-subblocks:without}]{\parbox{0.48\columnwidth}{\hfill}}
  \subfloat[$4 \times 4$ subblock processing\label{fig:mpa-4x4-subblocks:with}]{\parbox{0.48\columnwidth}{\hfill}}
  \caption{Resulting pixel shifts for MPA for an exemplary motion plane and motion vector. (a) Pixel shifts for full block processing, where the motion model needs to be calculated for each pixel position individually. (b) Pixel shifts for $4 \times 4$ subblock processing, where the motion model is calculated for only one pixel position per subblock (red arrow) and the resulting pixel shift is applied to all pixels within the same subblock (gray arrows).}
  \label{fig:mpa-4x4-subblocks}
\end{figure}

To reduce the computational complexity of MPA, the motion model is executed only once per $4 \times 4$ subblock similar to the realization of the newly introduced 4-parameter and 6-parameter affine motion models~\cite{Li2018, Zhang2019} as visualized in Fig.~\ref{fig:mpa-4x4-subblocks:with}.
This means that the motion model is executed for a specific input pixel position $\vec{p}_\text{o}$ within each subblock and the resulting pixel shift $\Delta \vec{p} = \vec{m}_\text{mpa}(\vec{p}_\text{o}, \vec{t}, \mat{R}) - \vec{p}_\text{o}$ is applied to all pixels within the same subblock.
This reduces the amount of calculations by a factor of $1/16$ compared to full block processing, where the resulting pixel shift is calculated for each pixel position individually as shown in Fig.~\ref{fig:mpa-4x4-subblocks:without}.

The precise pixel position within each subblock, at which the motion model is evaluated, can in principle be chosen arbitrarily.
While intuitively, one would choose the center position of the subblock, our experiments showed that better results are obtained using one of the actual pixel positions surrounding the floating point center position.
For further evaluations, we use the pixel position in the second row and second column of each subblock as shown in Fig.~\ref{fig:mpa-4x4-subblocks:with}.

After motion modeling, the reference frame sampling step shaded yellow in Fig.~\ref{fig:inter-prediction-pipeline} extracts the pixel values shifted according to the resulting motion field from the selected reference frame to generate a prediction for the current CU.
No adaptions for the integration of MPA are necessary here.

If no merge mode is applied, an encoder-side motion estimation needs to be performed in order to find suitable parameters for motion compensation.
While the search strategies for the translational and affine motion models remain unchanged, an extended strategy for the motion-plane-adaptive motion model is required.
Conceptually, the search strategy for MPA can be interpreted as three independent searches, one for each available motion plane.
Thereby, the existing motion vector search strategies for the classical translational motion estimation can be applied to search the best-matching motion vector on each motion plane.
Within MPA, the motion plane and motion vector combination leading to the minimum rate-distortion cost is then selected.
In case of bi-prediction, both predictions share the same motion plane.
In the end, the encoder decides between MPA and the classical pipeline by selecting the tool with the lowest overall rate-distortion cost.

\subsubsection*{Other Tools}
The H.266/VVC video coding standard features a number of additional inter prediction tools~\cite{Bross2021a}, all of which have been tested for compatibility to MPA or have been adapted accordingly where appropriate.
This includes the merge mode with motion vector difference (MMVD), the geometric partitioning mode (GPM), subblock temporal motion vector prediction (SbTMVP), decoder-side motion vector refinement (DMVR), bi-directional optical flow (BDOF), adaptive motion vector resolution (AMVR), bi-prediction with CU-level weights (BCW) and symmetric motion vector difference (SMVD).

The wrap-around reference frame padding introduced by the wrap-around motion compensation tool for 360-degree video is also compatible with MPA and leads to an improved interpolation quality for pixel values close to the boundaries of the image.

\begin{table}[t!]
  \centering
  \renewcommand{\arraystretch}{1.1}
  \caption{Overview over the test sequences used for the performance evaluation. All sequences are in erp projection format.}
  \label{table:test-sequences}
  \scalebox{1}{%
  \begin{tabular}{|l|c|c|c|c|}
    \hline
    Sequence & \makecell[l]{Bit\\depth} & \makecell[l]{Original\\resolution} & \makecell[l]{Coding\\resolution} & \makecell[l]{Color\\format} \\
    \hline
    SkateboardInLot & 10 & 4096 $\times$ 2048 & 2216 $\times$ 1108 & 4:0:0 \\
    \hline
    ChairliftRide & 10 & 4096 $\times$ 2048 & 2216 $\times$ 1108 & 4:0:0 \\
    \hline
    KiteFlite & 8 & 4096 $\times$ 2048 & 2216 $\times$ 1108 & 4:0:0 \\
    \hline
    Harbor & 8 & 4096 $\times$ 2048 & 2216 $\times$ 1108 & 4:0:0 \\
    \hline
    Trolley & 8 & 4096 $\times$ 2048 & 2216 $\times$ 1108 & 4:0:0 \\
    \hline
    GasLamp & 8 & 4096 $\times$ 2048 & 2216 $\times$ 1108 & 4:0:0 \\
    \hline
    Balboa & 8 & 3072 $\times$ 1536 & 2216 $\times$ 1108 & 4:0:0 \\
    \hline
    Broadway & 8 & 3072 $\times$ 1536 & 2216 $\times$ 1108 & 4:0:0 \\
    \hline
    Landing2 & 8 & 3072 $\times$ 1536 & 2216 $\times$ 1108 & 4:0:0 \\
    \hline
    BranCastle2 & 8 & 3072 $\times$ 1536 & 2216 $\times$ 1108 & 4:0:0 \\
    \hline
  \end{tabular}%
  }
\end{table}

\section{Performance Evaluation}\label{sec:performance}

\subsection{Experimental Setup}

We evaluate the performance of the proposed MPA inter prediction tool based on the introduced integration into the state-of-the-art H.266/VVC video coding standard.
As a baseline for our evaluations, the VVC reference software VTM-14.2~\cite{Browne2021,VTM-14.2} is employed.
The extended software including the proposed MPA tool as described in Subsection~\ref{subsec:coding} is termed MPA-VVC\footnote{The source code of our MPA-VVC implementation is publicly available at \textit{\{link will be provided upon publication\}}}, the baseline is termed VTM-14.2.

The employed 360-degree test sequences~\cite{Hanhart2018} are summarized in Table~\ref{table:test-sequences}.
All sequences are in ERP projection format.
In accordance with the common test conditions for 360-degree video (360-CTC)~\cite{Hanhart2018}, all tests are performed on 32 frames of each sequence using four quantization parameters $\text{QP} \in \{ 22, 27, 32, 37 \}$ with the random access (RA) configuration ~\cite{Bossen2020}.
For each sequence and QP value, three codecs are tested: VTM-14.2 as a baseline, MPA-VVC with MPA-MVP disabled, and MPA-VVC with MPA-MVP enabled.
Rate savings are calculated according to the Bjøntegaard Delta (BD) model with piecewise cubic interpolation~\cite{Bjontegaard2001}.
Complexity is reported as the encoding and decoding runtimes relative to the baseline, where all tests are performed on one core of an Intel\textsuperscript{\textregistered} Xeon\textsuperscript{\textregistered} E5-2690V3 with a base frequency of 2.60 GHz.

The 360-CTC specify to encode the sequences at a lower coding resolution (potentially performing a projection format conversion while down-sampling) and to up-sample the resulting sequences to the original resolution (and projection format) after decoding.
This allows a less biased comparison of different projection formats.
Quality metrics can then be calculated either at the original resolution (end-to-end) or at the coding resolution (codec).
Although we limit our investigations to the ERP format, we report the results for both configurations (end-to-end and codec).
Table~\ref{table:360-metrics} gives an overview over the employed quality metrics for 360-degree video that we evaluate using the 360Lib software \mbox{360Lib-12.0}~\cite{Ye2020a, 360Lib-12.0}.

\begin{table}
  \centering
  \renewcommand{\arraystretch}{1.1}
  \caption{Quality metrics according to the 360-CTC.}
  \label{table:360-metrics}
  \begin{tabular}{|p{0.23\linewidth}|p{0.6\linewidth}|}
    \hline
    PSNR\newline (codec) & Conventional Peak-Signal-to-Noise-Ratio as the fraction of maximum signal value over mean squared error (MSE). \\
    \hline
    WS-PSNR\newline (codec)\newline\newline E2E-WS-PSNR\newline (end-to-end) & Weighted-to-Spherically uniform PSNR, where the MSE in the PSNR calculation is replaced by a weighted MSE that weights the error at each pixel position by the spherical area covered by that pixel position~\cite{Ye2020a}.\\
    \hline
    S-PSNR-NN\newline (codec)\newline\newline E2E-S-PSNR-NN\newline (end-to-end) & Spherical PSNR, where the MSE is calculated for a set of uniformly sampled pixel positions on the sphere.
    The signal values are obtained at the corresponding pixel positions in the projection domain using nearest neighbor interpolation~\cite{Ye2020a}. \\
    \hline
    PSNR-DYN-VP0/1\newline (end-to-end) & Conventional PSNR, where the MSE is calculated for a predefined dynamic viewport (VP0 or VP1)~\cite{Hanhart2018}. \\
    \hline
  \end{tabular}
\end{table}

\begin{table*}[t]
  \centering
  \renewcommand{\arraystretch}{1.1}
  \caption{Bjøntegaard Delta rate savings in \% for MPA-VVC with MPA-MVP=OFF and ON with respect to the baseline VTM-14.2. Negative values (black) represent actual rate savings, positive values (red) represent increases in rate. Bold entries mark the highest rate savings among the tested codecs for each metric separately.}\label{table:bd-table}
  \begin{tabular}{l||r|r||r|r||r|r||r|r||r|r||r|r||r|r}
& \multicolumn{2}{c||}{PSNR} & \multicolumn{2}{c||}{WS-PSNR} & \multicolumn{2}{c||}{S-PSNR-NN} & \multicolumn{2}{c||}{\makecell[c]{E2E-\\WS-PSNR}} & \multicolumn{2}{c||}{\makecell[c]{E2E-\\S-PSNR-NN}} & \multicolumn{2}{c||}{\makecell[c]{PSNR-\\DYN-VP0}} & \multicolumn{2}{c}{\makecell[c]{PSNR-\\DYN-VP1}} \\ 
\hline 
\multicolumn{1}{r||}{MPA-MVP} & \multicolumn{1}{c|}{OFF}& \multicolumn{1}{c||}{ON}& \multicolumn{1}{c|}{OFF}& \multicolumn{1}{c||}{ON}& \multicolumn{1}{c|}{OFF}& \multicolumn{1}{c||}{ON}& \multicolumn{1}{c|}{OFF}& \multicolumn{1}{c||}{ON}& \multicolumn{1}{c|}{OFF}& \multicolumn{1}{c||}{ON}& \multicolumn{1}{c|}{OFF}& \multicolumn{1}{c||}{ON}& \multicolumn{1}{c|}{OFF} & \multicolumn{1}{c}{ON}\\ 
\hline 
SkateboardInLot & \textcolor{black}{-0.97} & \textbf{\textcolor{black}{-1.91}} & \textcolor{black}{-1.06} & \textbf{\textcolor{black}{-1.64}} & \textcolor{black}{-1.02} & \textbf{\textcolor{black}{-1.59}} & \textcolor{black}{-1.24} & \textbf{\textcolor{black}{-1.74}} & \textcolor{black}{-1.11} & \textbf{\textcolor{black}{-1.66}} & \textcolor{red}{+0.14} & \textbf{\textcolor{black}{-0.73}} & \textcolor{black}{-0.15} & \textbf{\textcolor{black}{-0.31}} \\ 
ChairliftRide & \textcolor{black}{-1.56} & \textbf{\textcolor{black}{-3.05}} & \textcolor{black}{-1.22} & \textbf{\textcolor{black}{-2.41}} & \textcolor{black}{-1.27} & \textbf{\textcolor{black}{-2.32}} & \textcolor{black}{-1.41} & \textbf{\textcolor{black}{-2.92}} & \textcolor{black}{-1.46} & \textbf{\textcolor{black}{-2.95}} & \textcolor{black}{-0.58} & \textbf{\textcolor{black}{-1.74}} & \textcolor{black}{-1.54} & \textbf{\textcolor{black}{-3.74}} \\ 
KiteFlite & \textbf{\textcolor{black}{-0.15}} & \textcolor{black}{-0.03} & \textcolor{red}{+0.03} & \textbf{\textcolor{black}{-0.02}} & \textcolor{red}{+0.03} & \textbf{\textcolor{black}{-0.05}} & \textcolor{red}{+0.02} & \textbf{\textcolor{black}{-0.13}} & \textcolor{red}{+0.03} & \textbf{\textcolor{black}{-0.13}} & \textcolor{red}{+0.10} & \textcolor{red}{+0.04} & \textcolor{red}{+0.15} & \textbf{\textcolor{black}{-0.33}} \\ 
Harbor & \textcolor{red}{+2.80} & \textcolor{red}{+1.40} & \textcolor{red}{+0.27} & \textbf{\textcolor{black}{-0.15}} & \textcolor{red}{+0.17} & \textbf{\textcolor{black}{-0.17}} & \textcolor{red}{+0.36} & \textbf{\textcolor{black}{-0.18}} & \textcolor{red}{+0.21} & \textbf{\textcolor{black}{-0.28}} & \textcolor{red}{+0.01} & \textbf{\textcolor{black}{-0.23}} & \textcolor{black}{-0.11} & \textbf{\textcolor{black}{-0.28}} \\ 
Trolley & \textcolor{red}{+0.11} & \textbf{\textcolor{black}{-0.38}} & \textcolor{red}{+0.04} & \textcolor{red}{+0.01} & \textcolor{red}{+0.04} & \textcolor{red}{+0.01} & \textbf{\textcolor{black}{-0.11}} & \textcolor{black}{-0.10} & \textcolor{black}{-0.13} & \textbf{\textcolor{black}{-0.13}} & \textcolor{red}{+0.06} & \textcolor{red}{+0.20} & \textcolor{red}{+0.17} & \textcolor{red}{+0.24} \\ 
GasLamp & \textcolor{red}{+1.18} & \textbf{\textcolor{black}{-1.27}} & \textcolor{black}{-0.08} & \textbf{\textcolor{black}{-0.43}} & \textcolor{black}{-0.18} & \textbf{\textcolor{black}{-0.46}} & \textcolor{black}{-0.02} & \textbf{\textcolor{black}{-0.43}} & \textcolor{black}{-0.10} & \textbf{\textcolor{black}{-0.49}} & \textcolor{red}{+0.01} & \textbf{\textcolor{black}{-0.12}} & \textcolor{black}{-0.07} & \textbf{\textcolor{black}{-0.47}} \\ 
Balboa & \textcolor{black}{-2.05} & \textbf{\textcolor{black}{-3.46}} & \textcolor{black}{-1.87} & \textbf{\textcolor{black}{-3.24}} & \textcolor{black}{-1.84} & \textbf{\textcolor{black}{-3.19}} & \textcolor{black}{-2.13} & \textbf{\textcolor{black}{-3.52}} & \textcolor{black}{-2.05} & \textbf{\textcolor{black}{-3.52}} & \textcolor{black}{-1.79} & \textbf{\textcolor{black}{-3.49}} & \textcolor{black}{-2.65} & \textbf{\textcolor{black}{-4.09}} \\ 
Broadway & \textcolor{black}{-1.63} & \textbf{\textcolor{black}{-3.12}} & \textcolor{black}{-1.64} & \textbf{\textcolor{black}{-3.11}} & \textcolor{black}{-1.61} & \textbf{\textcolor{black}{-3.09}} & \textcolor{black}{-1.98} & \textbf{\textcolor{black}{-3.67}} & \textcolor{black}{-1.98} & \textbf{\textcolor{black}{-3.55}} & \textcolor{black}{-2.37} & \textbf{\textcolor{black}{-3.95}} & \textcolor{black}{-3.19} & \textbf{\textcolor{black}{-5.40}} \\ 
Landing2 & \textcolor{black}{-3.37} & \textbf{\textcolor{black}{-3.96}} & \textcolor{black}{-2.92} & \textbf{\textcolor{black}{-3.40}} & \textcolor{black}{-2.89} & \textbf{\textcolor{black}{-3.42}} & \textcolor{black}{-3.45} & \textbf{\textcolor{black}{-3.94}} & \textcolor{black}{-3.38} & \textbf{\textcolor{black}{-3.87}} & \textcolor{black}{-3.07} & \textbf{\textcolor{black}{-3.47}} & \textbf{\textcolor{black}{-4.01}} & \textcolor{black}{-3.46} \\ 
BranCastle2 & \textcolor{black}{-0.74} & \textbf{\textcolor{black}{-1.42}} & \textcolor{black}{-0.58} & \textbf{\textcolor{black}{-1.19}} & \textcolor{black}{-0.62} & \textbf{\textcolor{black}{-1.19}} & \textcolor{black}{-0.72} & \textbf{\textcolor{black}{-1.39}} & \textcolor{black}{-0.72} & \textbf{\textcolor{black}{-1.37}} & \textcolor{black}{-0.89} & \textbf{\textcolor{black}{-1.11}} & \textcolor{black}{-0.80} & \textbf{\textcolor{black}{-1.28}} \\ 
\hline 
Average & -0.64 & \textbf{-1.72} & -0.90 & \textbf{-1.56} & -0.92 & \textbf{-1.55} & -1.07 & \textbf{-1.80} & -1.07 & \textbf{-1.80} & -0.84 & \textbf{-1.46} & -1.22 & \textbf{-1.91} \\ 
\end{tabular}
  \vspace*{-8pt}
\end{table*}

\begin{figure*}[t]
\setlength\figurewidth{0.34\linewidth}
\setlength\figureheight{.23\linewidth}
\centering
{
  \footnotesize%
%
%

\begin{tikzpicture}

    \definecolor{color0}{rgb}{0.172549019607843,0.627450980392157,0.172549019607843}
    \definecolor{color1}{rgb}{1,0.498039215686275,0.0549019607843137}
    \definecolor{color2}{rgb}{0.83921568627451,0.152941176470588,0.156862745098039}

    \begin{axis}[%
    hide axis,
    xmin=10,
    xmax=50,
    ymin=0,
    ymax=0.4,
    legend style={draw=white!15!black,legend cell align=left},
    legend columns=6
    ]
    \addlegendimage{semithick, color0, mark=square*, mark size=1.25, mark options={solid}};
    \addlegendentry{};
    \addlegendimage{semithick, color0, dashed, mark=square, mark size=1.25, mark options={solid,fill=none}};
    \addlegendentry{VTM-14.2\phantom{MM}};
    \addlegendimage{semithick, color1, mark=triangle*, mark size=1.75, mark options={solid}};
    \addlegendentry{};
    \addlegendimage{semithick, color1, dashed, mark=triangle, mark size=1.75, mark options={solid,fill=none}};
    \addlegendentry{MPA-VVC (MPA-MVP=OFF)\phantom{MM}};
    \addlegendimage{semithick, color2, mark=*, mark size=1, mark options={solid}};
    \addlegendentry{};
    \addlegendimage{semithick, color2, dashed, mark=o, mark size=1, mark options={solid,fill=none}};
    \addlegendentry{MPA-VVC (MPA-MVP=ON)\phantom{MM}};
    \addlegendimage{only marks, mark size=0};
    \addlegendentry{};
    \addlegendimage{semithick, black};
    \addlegendentry{PSNR};
    \addlegendimage{only marks, mark size=0};Z
    \addlegendentry{};
    \addlegendimage{semithick, black, dashed};
    \addlegendentry{WS-PSNR};
    \end{axis}
\end{tikzpicture}\\[-134pt]
  \subfloat[\textit{Landing2}\label{fig:rd-plots:landing2}]{%
\begin{tikzpicture}

\definecolor{color0}{rgb}{0.172549019607843,0.627450980392157,0.172549019607843}
\definecolor{color1}{rgb}{1,0.498039215686275,0.0549019607843137}
\definecolor{color2}{rgb}{0.83921568627451,0.152941176470588,0.156862745098039}

\begin{axis}[
height=\figureheight,
legend style={fill opacity=0.8, draw opacity=1, text opacity=1, draw=white!80!black},
tick align=outside,
tick pos=left,
width=\figurewidth,
x grid style={white!69.0196078431373!black},
xlabel={Average bits per pixel},
xmajorgrids,
xmin=0.171090787055742, xmax=6.91120534608818,
xtick style={color=black},
y grid style={white!69.0196078431373!black},
ylabel={WS-PSNR/PSNR in dB},
ymajorgrids,
ymin=31.6584603125, ymax=41.4189459375,
ytick style={color=black}
]
\addplot [thick, color0, forget plot]
table {%
0.501710565757406 32.385034375
0.55178236204369 32.8309083029789
0.640831713846797 33.2890316589367
0.760302686828009 33.758974613405
0.90163934664861 34.2403073369156
1.05628575896988 34.7326
1.23532101334574 35.2288957179494
1.45870895656653 35.7301974261899
1.73122079014817 36.2480150254557
2.05762771560656 36.793858416481
2.44270093445764 37.3792375
2.96420109934932 38.0120541555401
3.67449073259094 38.6902793156076
4.54024384019054 39.4105264604051
5.52813442815615 40.169409070135
6.6048365024958 40.963540625
};
\addplot [semithick, color0, mark=square*, mark size=1.25, mark options={solid}, only marks, forget plot]
table {%
6.6048365024958 40.963540625
2.44270093445764 37.3792375
1.05628575896988 34.7326
0.501710565757406 32.385034375
};
\addplot [thick, color1, forget plot]
table {%
0.47861304070169 32.43283125
0.52332516157083 32.8728197862334
0.607784526925996 33.3254177337
0.723021588314159 33.79016641305
0.860066797282292 34.2666071449334
1.00995060537737 34.75428125
1.18497556968498 35.2462366993056
1.40607063659553 35.7435437446667
1.67720584161319 36.2577771278751
2.00235122024214 36.8005115907223
2.38547680798655 37.383321875
2.90493444329879 38.0144454730443
3.61404303473728 38.6920498134249
4.4792232229653 39.4126045422833
5.46689564864612 40.1725793057611
6.54348095244301 40.96844375
};
\addplot [semithick, color1, mark=triangle*, mark size=1.75, mark options={solid}, only marks, forget plot]
table {%
6.54348095244301 40.96844375
2.38547680798655 37.383321875
1.00995060537737 34.75428125
0.47861304070169 32.43283125
};
\addplot [thick, color2, forget plot]
table {%
0.477459630648125 32.4503375
0.523156243768964 32.8917002625318
0.608177303033358 33.3452886625953
0.723706348664387 33.8106706813929
0.860926920885133 34.287414300127
1.01102255991867 34.7750875
1.18590198238473 35.2665599932249
1.40613692776276 35.7628398236629
1.67590547056552 36.2757024699884
1.99938568530572 36.816923410876
2.38075564649611 37.398278125
2.89805377789399 38.0279836265923
3.60423642774612 38.7040588924163
4.46585855729934 39.4229823449442
5.44947512780053 40.1812324066481
6.52164110049655 40.9752875
};
\addplot [semithick, color2, mark=*, mark size=1, mark options={solid}, only marks, forget plot]
table {%
6.52164110049655 40.9752875
2.38075564649611 37.398278125
1.01102255991867 34.7750875
0.477459630648125 32.4503375
};
\addplot [thick, color0, dashed, forget plot]
table {%
0.501710565757406 32.10211875
0.55178236204369 32.5190369317863
0.640831713846797 32.9488325453589
0.760302686828009 33.3910065680384
0.90163934664861 33.8450599771452
1.05628575896988 34.31049375
1.23532101334574 34.7799397641549
1.45870895656653 35.2545624058619
1.73122079014817 35.7466614154915
2.05762771560656 36.2685365334141
2.44270093445764 36.8324875
2.96420109934932 37.4480841597996
3.67449073259094 38.1139622891496
4.54024384019054 38.8260205260997
5.52813442815615 39.5801575086999
6.6048365024958 40.372271875
};
\addplot [semithick, color0, mark=square, mark size=1.25, mark options={solid,fill=none}, only marks, forget plot]
table {%
6.6048365024958 40.372271875
2.44270093445764 36.8324875
1.05628575896988 34.31049375
0.501710565757406 32.10211875
};
\addplot [thick, color1, dashed, forget plot]
table {%
0.47861304070169 32.106078125
0.52332516157083 32.5225472908582
0.607784526925996 32.9518477475746
0.723021588314159 33.3934833088619
0.860066797282292 33.8469577884328
1.00995060537737 34.311775
1.18497556968498 34.7804689168432
1.40607063659553 35.2541973069662
1.67720584161319 35.7453996761676
2.00235122024214 36.266515530246
2.38547680798655 36.829984375
2.90493444329879 37.4455084038958
3.61404303473728 38.1117247512578
4.4792232229653 38.8244753341718
5.46689564864612 39.5796020697239
6.54348095244301 40.372946875
};
\addplot [semithick, color1, mark=triangle, mark size=1.75, mark options={solid,fill=none}, only marks, forget plot]
table {%
6.54348095244301 40.372946875
2.38547680798655 36.829984375
1.00995060537737 34.311775
0.47861304070169 32.106078125
};
\addplot [thick, color2, dashed, forget plot]
table {%
0.477459630648125 32.11885
0.523156243768964 32.5357603695142
0.608177303033358 32.9653589835427
0.723706348664387 33.407160225314
0.860926920885133 33.8606784780569
1.01102255991867 34.325428125
1.18590198238473 34.793881087798
1.40613692776276 35.2671704722508
1.67590547056552 35.7578024378045
1.99938568530572 36.2782831439055
2.38075564649611 36.84111875
2.89805377789399 37.4559964365803
3.60423642774612 38.1214981786528
4.46585855729934 38.8334723274352
5.44947512780053 39.5877672341451
6.52164110049655 40.38023125
};
\addplot [semithick, color2, mark=o, mark size=1, mark options={solid,fill=none}, only marks, forget plot]
table {%
6.52164110049655 40.38023125
2.38075564649611 36.84111875
1.01102255991867 34.325428125
0.477459630648125 32.11885
};
\end{axis}

\end{tikzpicture}}%
  \hfill%
  \subfloat[\textit{Balboa}\label{fig:rd-plots:balboa}]{%
\begin{tikzpicture}

\definecolor{color0}{rgb}{0.172549019607843,0.627450980392157,0.172549019607843}
\definecolor{color1}{rgb}{1,0.498039215686275,0.0549019607843137}
\definecolor{color2}{rgb}{0.83921568627451,0.152941176470588,0.156862745098039}

\begin{axis}[
height=\figureheight,
legend style={fill opacity=0.8, draw opacity=1, text opacity=1, draw=white!80!black},
tick align=outside,
tick pos=left,
width=\figurewidth,
x grid style={white!69.0196078431373!black},
xlabel={Average bits per pixel},
xmajorgrids,
xmin=0.096825352865279, xmax=4.28519122496057,
xtick style={color=black},
y grid style={white!69.0196078431373!black},
ymajorgrids,
ymin=31.71137296875, ymax=42.79955515625,
ytick style={color=black}
]
\addplot [thick, color0, forget plot]
table {%
0.291050319957252 33.5377125
0.323701741286682 33.9408681745541
0.384155700211882 34.3701197736622
0.466192366410826 34.8236591604933
0.563591909561487 35.2996781982163
0.670134499341839 35.79636875
0.796340971451731 36.31812496231
0.956724263309222 36.8691658210854
1.15105269716961 37.4474204487059
1.37909459528818 38.050817967551
1.64061827992024 38.6772875
1.97162071564789 39.330892232895
2.39796847185961 40.0159150276318
2.90423427802018 40.7306145184212
3.47499086359443 41.4732493394737
4.09481095804715 42.242078125
};
\addplot [semithick, color0, mark=square*, mark size=1.25, mark options={solid}, only marks, forget plot]
table {%
4.09481095804715 42.242078125
1.64061827992024 38.6772875
0.670134499341839 35.79636875
0.291050319957252 33.5377125
};
\addplot [thick, color1, forget plot]
table {%
0.288740241629632 33.56608125
0.321241703890886 33.9732166944471
0.381000327041032 34.4053558333412
0.461937352764888 34.8608313125118
0.557974022747268 35.3379757777882
0.663031578672992 35.835121875
0.787192593375912 36.3557425134141
0.944640836258123 36.9037281860952
1.1354493438568 37.4780379145692
1.35969115270912 38.0776307203622
1.61743929935227 38.701465625
1.94522906392548 39.3538021851907
2.36910838407486 40.0383769270896
2.87341928259789 40.753364326393
3.44250378229202 41.4969388587977
4.06070390595472 42.267275
};
\addplot [semithick, color1, mark=triangle*, mark size=1.75, mark options={solid}, only marks, forget plot]
table {%
4.06070390595472 42.267275
1.61743929935227 38.701465625
0.663031578672992 35.835121875
0.288740241629632 33.56608125
};
\addplot [thick, color2, forget plot]
table {%
0.287205619778702 33.59560625
0.32011983589536 34.0027376512787
0.379917692987595 34.434925578836
0.460636747486189 34.8904965557539
0.556314555821922 35.3677771051146
0.660988674425576 35.86509375
0.784446691321386 36.3860165582088
0.940600974837531 36.9344503092163
1.1296762518607 37.5092235936194
1.35189724927759 38.109165002015
1.6074886939749 38.733103125
1.93280534181032 39.3852375717064
2.35366505761492 40.0693490025447
2.85448723173537 40.7836375850298
3.41969125451838 41.5263034866766
4.03369651631065 42.295546875
};
\addplot [semithick, color2, mark=*, mark size=1, mark options={solid}, only marks, forget plot]
table {%
4.03369651631065 42.295546875
1.6074886939749 38.733103125
0.660988674425576 35.86509375
0.287205619778702 33.59560625
};
\addplot [thick, color0, dashed, forget plot]
table {%
0.291050319957252 32.21538125
0.323701741286682 32.6162263057564
0.384155700211882 33.0452672922692
0.466192366410826 33.5004307509038
0.563591909561487 33.9796432230256
0.670134499341839 34.48083125
0.796340971451731 35.0086892018107
0.956724263309222 35.5680551486052
1.15105269716961 36.1570711819944
1.37909459528818 36.773879393589
1.64061827992024 37.416621875
1.97162071564789 38.0902758006546
2.39796847185961 38.7996349007364
2.90423427802018 39.5425661004909
3.47499086359443 40.3169363251636
4.09481095804715 41.1206125
};
\addplot [semithick, color0, mark=square, mark size=1.25, mark options={solid,fill=none}, only marks, forget plot]
table {%
4.09481095804715 41.1206125
1.64061827992024 37.416621875
0.670134499341839 34.48083125
0.291050319957252 32.21538125
};
\addplot [thick, color1, dashed, forget plot]
table {%
0.288740241629632 32.23355
0.321241703890886 32.6393769813411
0.381000327041032 33.0721836940233
0.461937352764888 33.530064416035
0.557974022747268 34.0111134253645
0.663031578672992 34.513425
0.787192593375912 35.0406271110587
0.944640836258123 35.5971664432345
1.1354493438568 36.182365344881
1.35969115270912 36.7955461643516
1.61743929935227 37.43603125
1.94522906392548 38.1090549793073
2.36910838407486 38.8188079610957
2.87341928259789 39.5630468907304
3.44250378229202 40.3395284635768
4.06070390595472 41.146009375
};
\addplot [semithick, color1, mark=triangle, mark size=1.75, mark options={solid,fill=none}, only marks, forget plot]
table {%
4.06070390595472 41.146009375
1.61743929935227 37.43603125
0.663031578672992 34.513425
0.288740241629632 32.23355
};
\addplot [thick, color2, dashed, forget plot]
table {%
0.287205619778702 32.2648125
0.32011983589536 32.669485845851
0.379917692987595 33.101423412553
0.460636747486189 33.5586808688294
0.556314555821922 34.0393138834039
0.660988674425576 34.541378125
0.784446691321386 35.0687865139843
0.940600974837531 35.6260758233352
1.1296762518607 36.2122375006941
1.35189724927759 36.8262629937019
1.6074886939749 37.46714375
1.93280534181032 38.1400163604471
2.35366505761492 38.849222186753
2.85448723173537 39.5925598328353
3.41969125451838 40.3678279026118
4.03369651631065 41.172825
};
\addplot [semithick, color2, mark=o, mark size=1, mark options={solid,fill=none}, only marks, forget plot]
table {%
4.03369651631065 41.172825
1.6074886939749 37.46714375
0.660988674425576 34.541378125
0.287205619778702 32.2648125
};
\end{axis}

\end{tikzpicture}}%
  \hfill%
  \subfloat[\textit{Harbor}\label{fig:rd-plots:harbor}]{%
\begin{tikzpicture}

\definecolor{color0}{rgb}{0.172549019607843,0.627450980392157,0.172549019607843}
\definecolor{color1}{rgb}{1,0.498039215686275,0.0549019607843137}
\definecolor{color2}{rgb}{0.83921568627451,0.152941176470588,0.156862745098039}

\begin{axis}[
height=\figureheight,
legend style={fill opacity=0.8, draw opacity=1, text opacity=1, draw=white!80!black},
tick align=outside,
tick pos=left,
width=\figurewidth,
x grid style={white!69.0196078431373!black},
xlabel={Average bits per pixel},
xmajorgrids,
xmin=0.0908416309348486, xmax=1.75908799150256,
xtick style={color=black},
y grid style={white!69.0196078431373!black},
ymajorgrids,
ymin=33.69022765625, ymax=43.89926921875,
ytick style={color=black}
]
\addplot [thick, color0, forget plot]
table {%
0.167133678270276 34.81728125
0.195333468366219 35.3677508542878
0.232723938573614 35.9045018128635
0.277744237826927 36.4269185942952
0.328833515060622 36.9343856671513
0.384430919209165 37.4262875
0.446700542858989 37.8900000406905
0.519284371829676 38.3269595292302
0.602838394905412 38.7576408724247
0.698018600870383 39.2025189770794
0.805480978508778 39.68206875
0.934219487801806 40.2001704936327
1.09022106002278 40.7459459147118
1.26963114089404 41.3177320264745
1.46859517613792 41.9138658421582
1.68325861147676 42.532684375
};
\addplot [semithick, color0, mark=square*, mark size=1.25, mark options={solid}, only marks, forget plot]
table {%
1.68325861147676 42.532684375
0.805480978508778 39.68206875
0.384430919209165 37.4262875
0.167133678270276 34.81728125
};
\addplot [thick, color1, forget plot]
table {%
0.168026430684617 34.74665625
0.195639409591114 35.2766145709864
0.232555862523293 35.7972697129592
0.277175946035066 36.3083444444388
0.327899816680346 36.8095615339456
0.383127631013046 37.30064375
0.445155931335995 37.7717183298617
0.517717685690066 38.223432146983
0.601367689236231 38.6712874241735
0.696660737135463 39.1307863842427
0.804151624548736 39.61743125
0.932802643613451 40.1335696947709
1.08864556812996 40.6705438128673
1.26783941575366 41.2273474585782
1.46654320413995 41.8029744861927
1.68091595094423 42.39641875
};
\addplot [semithick, color1, mark=triangle*, mark size=1.75, mark options={solid}, only marks, forget plot]
table {%
1.68091595094423 42.39641875
0.804151624548736 39.61743125
0.383127631013046 37.30064375
0.168026430684617 34.74665625
};
\addplot [thick, color2, forget plot]
table {%
0.166671010960654 34.801425
0.195387572171804 35.3250132538781
0.233112631150037 35.8405106366342
0.278338484970047 36.3477082674513
0.329557430706533 36.8463972655123
0.38526176543419 37.33636875
0.447426833042783 37.8077440543125
0.519570607024685 38.261010567023
0.60249932885851 38.7115084900773
0.697019240022869 39.174578025421
0.803936581996377 39.665559375
0.932353734665054 40.1870488207009
1.08823490362841 40.7306740482886
1.26767906893529 41.2953223655257
1.46678521063455 41.8798810801752
1.68165230877504 42.4832375
};
\addplot [semithick, color2, mark=*, mark size=1, mark options={solid}, only marks, forget plot]
table {%
1.68165230877504 42.4832375
0.803936581996377 39.665559375
0.38526176543419 37.33636875
0.166671010960654 34.801425
};
\addplot [thick, color0, dashed, forget plot]
table {%
0.167133678270276 34.18060625
0.195333468366219 34.7454409583937
0.232723938573614 35.3143593751811
0.277744237826927 35.8873191877717
0.328833515060622 36.4642780835748
0.384430919209165 37.04519375
0.446700542858989 37.6252296938692
0.519284371829676 38.2044846445605
0.602838394905412 38.7903191233171
0.698018600870383 39.3900936513825
0.805480978508778 40.01116875
0.934219487801806 40.6552477902239
1.09022106002278 41.3186404515018
1.26963114089404 42.0006167176679
1.46859517613792 42.700446572556
1.68325861147676 43.4174
};
\addplot [semithick, color0, mark=square, mark size=1.25, mark options={solid,fill=none}, only marks, forget plot]
table {%
1.68325861147676 43.4174
0.805480978508778 40.01116875
0.384430919209165 37.04519375
0.167133678270276 34.18060625
};
\addplot [thick, color1, dashed, forget plot]
table {%
0.168026430684617 34.154275
0.195639409591114 34.7233816286882
0.232555862523293 35.2958276360645
0.277175946035066 35.8715847040968
0.327899816680346 36.4506245147527
0.383127631013046 37.03291875
0.445155931335995 37.6132714489008
0.517717685690066 38.1917486868638
0.601367689236231 38.7762152003762
0.696660737135463 39.3745357259258
0.804151624548736 39.994575
0.932802643613451 40.638056820386
1.08864556812996 41.3009693135592
1.26783941575366 41.9825737090395
1.46654320413995 42.6821312363465
1.68091595094423 43.398903125
};
\addplot [semithick, color1, mark=triangle, mark size=1.75, mark options={solid,fill=none}, only marks, forget plot]
table {%
1.68091595094423 43.398903125
0.804151624548736 39.994575
0.383127631013046 37.03291875
0.168026430684617 34.154275
};
\addplot [thick, color2, dashed, forget plot]
table {%
0.166671010960654 34.181984375
0.195387572171804 34.7468025212119
0.233112631150037 35.3157549386357
0.278338484970047 35.8887982829536
0.329557430706533 36.4658892098477
0.38526176543419 37.046984375
0.447426833042783 37.6270934613586
0.519570607024685 38.206317605665
0.60249932885851 38.7922506442922
0.697019240022869 39.3924864136129
0.803936581996377 40.01461875
0.932353734665054 40.6604462901752
1.08823490362841 41.3261912170721
1.26767906893529 42.0110826863814
1.46678521063455 42.7143498537938
1.68165230877504 43.435221875
};
\addplot [semithick, color2, mark=o, mark size=1, mark options={solid,fill=none}, only marks, forget plot]
table {%
1.68165230877504 43.435221875
0.803936581996377 40.01461875
0.38526176543419 37.046984375
0.166671010960654 34.181984375
};
\end{axis}

\end{tikzpicture}}%
\hfill
}
\caption{Average rate-distortion curves for the sequences (a) \textit{Landing2}, (b) \textit{Balboa} and (c) \textit{Harbor} based on PSNR (solid) and WS-PSNR (dashed) for VTM-14.2 (green), MPA-VVC with MPA-MVP=OFF (orange), and MPA-VVC with MPA-MVP=ON (red). Best to be viewed enlarged on a monitor.}
\label{fig:rd-plots}
\end{figure*}
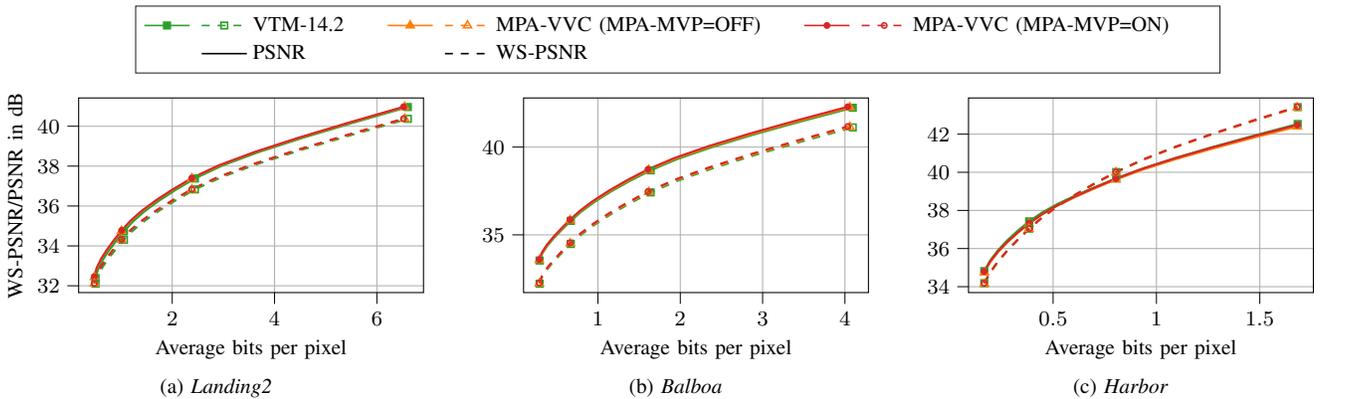

The focal length that is required by the generalized perspective projection is calculated as
\begin{align}
  f = \frac{1}{\tan (\pi/V)},
\end{align}
which is derived based on~\eqref{eq:perspective:radius} by requiring that for pixels at the optical center of the regarded motion planes, a shift by one pixel in the ERP domain (i.e., a shift of the incident angle by $\pi/V$), also leads to a shift by one pixel on the regarded motion plane.
This prevents undesired scaling when applying MPA.

\subsection{Rate-Distortion Analysis}

Table~\ref{table:bd-table} displays the rate savings that are achieved by MPA-VVC with respect to the baseline VTM-14.2 for each test sequence and quality metric.
For each quality metric, the left column shows the results if MPA-MVP is disabled (MPA-MVP=OFF), while the right column shows the results if MPA-MVP is enabled (MPA-MVP=ON).
Negative values (black) represent actual rate savings with respect to the baseline, positive values (red) represent increases in rate.
Bold entries represent the best compression efficiency (lowest BD-rate) among the tested codecs.

Our proposed MPA-VVC achieves significant rate savings across all quality metrics and in both MPA-MVP configurations.
For MPA-MVP=OFF, average rate savings of 0.64\% based on PSNR, 0.90\% based on WS-PSNR and 0.92\% based on S-PSNR-NN are achieved.
With the introduction of MPA-MVP, average rate savings considerably increase to 1.72\% based on PSNR, 1.56\% based on WS-PSNR and 1.55\% based on S-PSNR-NN.

With respect to end-to-end quality metrics, average rate savings of 1.07\% and 1.80\% are achieved with MPA-MVP=OFF and ON, respectively.
The BD-rate results for the two dynamic viewports support the positive impression with average rate savings of 0.84\% and 1.22\% for MPA-MVP=OFF, and 1.46\% and 1.91\% for MPA-MVP=ON.

Remarkably, the highest overall rate savings with 3.96\% based on PSNR, 3.40\% based on WS-PSNR and 3.42\% based on S-PSNR-NN for MPA-MVP=ON are achieved for the sequence \textit{Landing2}, which features rich rotational and translational camera motion that is not aligned with the ground surface.
This shows that the motion planes of MPA are able to better replicate some of the motion characteristics of 360-degree video than the classical translational and affine motion models despite the camera motion not being aligned to the available motion planes.
Fig.~\ref{fig:rd-plots:landing2} shows the corresponding rate-distortion curves based on PSNR (solid) and WS-PSNR (dashed), where it is visible that for higher QP values, MPA-VVC achieves coding gains mainly through an improved quality, while for lower QP values, MPA-VVC achieves coding gains mainly through a reduced rate.

As expected, significant rate savings are also achieved for the sequences \textit{SkateboardInLot}, \textit{ChairliftRide}, \textit{Balboa} and \textit{Broadway}, which feature global camera motion with little to no camera rotation.
Here, some of the motion surfaces in the videos are aligned with one of the available motion planes in MPA, such that the assumption of three exclusive motion planes is more likely to be satisfied.
Fig.~\ref{fig:rd-plots:balboa} shows the corresponding rate-distortion curves for the sequence \textit{Balboa}.
Similar to the sequence \textit{Landing2}, coding gains are achieved mainly through an improved quality for higher QP values and a reduced rate for lower QP values.

Neglectable rate savings or even slight increases in rate can be observed based on some metrics for the sequences \textit{KiteFlite} and \textit{Trolley} for both MPA-MVP=OFF and ON.
The sequences have in common that the camera is static, i.e., there is no global camera motion, and only little, locally constrained object motion, such that high compression rates are achieved.
As a result of this, small overheads in signaling such as the additional side information required for MPA, have a bigger effect on the overall rate.

A notable increase in rate based on PSNR can be observed for the sequence \textit{Harbor}, which, similar to the sequences \textit{KiteFlite} and \textit{Trolley}, features no global camera motion and only little object motion.
Fig.~\ref{fig:rd-plots:harbor} shows the corresponding rate-distortion curves.
Slight differences between the VTM-14.2 and MPA-VVC are visible for the PSNR-based rate-distortion points at QP=32 and QP=22.
However, based on WS-PSNR, the differences diminish and indeed small rate savings of 0.15\% can be observed for MPA-MVP=ON on this sequence, as well.

\subsection{Visual Examples}

\begin{figure*}[t!]
  \centering
  \newlength{\framewidth}
  \setlength{\framewidth}{0.14\linewidth}
  \subfloat[\textit{Broadway}, QP32, Frame 16, Wall, Original\label{fig:mpa-preds:1}]{%
  \includegraphics[width=\framewidth]{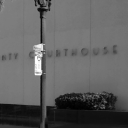}}
  \hfill
  \subfloat[\textit{Broadway}, QP32, Frame 16, Wall, VTM-14.2\label{fig:mpa-preds:2}]{%
  \includegraphics[width=\framewidth]{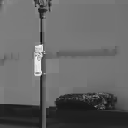}}
  \hfill
  \subfloat[\textit{Broadway}, QP32, Frame 16, Wall, MPA\label{fig:mpa-preds:3}]{%
  \includegraphics[width=\framewidth]{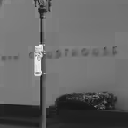}}
  \hfill
  \subfloat[\textit{Broadway}, QP32, Frame 16, Midlane, Original\label{fig:mpa-preds:4}]{%
  \includegraphics[width=\framewidth]{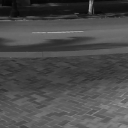}}
  \hfill
  \subfloat[\textit{Broadway}, QP32, Frame 16, Midlane, VTM-14.2\label{fig:mpa-preds:5}]{%
  \includegraphics[width=\framewidth]{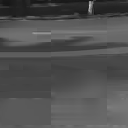}}
  \hfill
  \subfloat[\textit{Broadway}, QP32, Frame 16, Midlane, MPA\label{fig:mpa-preds:6}]{%
  \includegraphics[width=\framewidth]{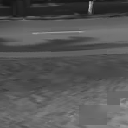}}
  \\
  \subfloat[\textit{Landing2}, QP32, Frame 4, Field, Original\label{fig:mpa-preds:7}]{%
  \includegraphics[width=\framewidth]{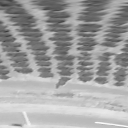}}
  \hfill
  \subfloat[\textit{Landing2}, QP32, Frame 4, Field, VTM-14.2\label{fig:mpa-preds:8}]{%
  \includegraphics[width=\framewidth]{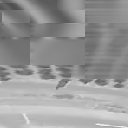}}
  \hfill
  \subfloat[\textit{Landing2}, QP32, Frame 4, Field, MPA\label{fig:mpa-preds:9}]{%
  \includegraphics[width=\framewidth]{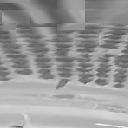}}
  \hfill
  \subfloat[\textit{SkateboardInLot}, Frame 6, QP27, Street, Original\label{fig:mpa-preds:10}]{%
  \includegraphics[width=\framewidth]{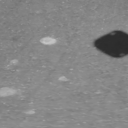}}
  \hfill
  \subfloat[\textit{SkateboardInLot}, Frame 6, QP27, Street, VTM-14.2\label{fig:mpa-preds:11}]{%
  \includegraphics[width=\framewidth]{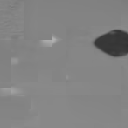}}
  \hfill
  \subfloat[\textit{SkateboardInLot}, Frame 6, QP27, Street, MPA\label{fig:mpa-preds:12}]{%
  \includegraphics[width=\framewidth]{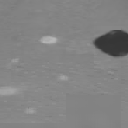}}
  \caption{Crops from original and predicted frames for VTM-14.2 and the proposed MPA with MPA-MVP=ON. Best to be viewed enlarged on a monitor.}
  \label{fig:mpa-preds}
\end{figure*}

Fig.~\ref{fig:mpa-preds} presents sections of the inter predicted frames for VTM-14.2 and the proposed MPA with MPA-MVP=ON.
For reference, the according sections from the original frames are shown as well.
It is clearly visible that MPA is able to retain textures at a higher level of detail with less distracting blocking artifacts than VTM-14.2 is able to.

In Fig.~\ref{fig:mpa-preds:1} - (c), the text appears completely blurred in the inter predicted frame for VTM-14.2, while most letters are still readable for MPA.
In Fig.~\ref{fig:mpa-preds:4} - (f), the tile texture of the midlane surface is heavily blurred and shows notable blocking artifacts for VTM-14.2, whereas MPA retains a much more detailed texture and shows less blocking artifacts.
In Fig.~\ref{fig:mpa-preds:7} - (i), the crop field close to the north pole of the frame shows strong blocking artifacts with notable blur, while MPA is able to retain a higher level of detail.
In Fig.~\ref{fig:mpa-preds:10} - (l), the street texture as well as the white marks are barely visible for VTM-14.2, whereas MPA preserves these marks and shows a finer street texture.
The clearly visible improvements in the quality of the inter predicted frames obtained by MPA explain the achieved rate savings.

\subsection{Motion Model Assessment}

Table~\ref{table:models-table} shows the rate savings that are achieved by different 360-degree motion models with respect to VTM-14.2 based on WS-PSNR.
The comparative motion models are abbreviated as 3DT for the 3D-translational motion model~\cite{Li2017, Li2019}, TAN for the tangential motion model~\cite{DeSimone2017}, ROT for the rotational motion model~\cite{Vishwanath2017, Vishwanath2018a}, GED for the geodesic motion model~\cite{Vishwanath2018}, and TAN+ROT for the combined tangential and rotational motion model~\cite{Marie2021}.
All motion models are using the same tools and extensions as MPA as described in Section~\ref{subsec:coding} with the exception of the adapted motion vector prediction described in Section~\ref{subsec:mpamvp} that is not available for the comparative motion models.
To ensure a fair comparison, MPA-MVP is thus disabled for MPA as well.
The global camera motion vector required by GED is provided as prior information.

Please note that most comparative 360-degree motion models have originally been introduced in the context of the HEVC video coding standard~\cite{Sullivan2012}, which did not feature the 4- and 6-parameter affine motion models available in VVC.
The available affine motion models could be one of the main reasons for the reduced rate savings of the comparative 360-degree motion models with respect to their original publications.

Overall, MPA shows the highest average rate savings of 0.90\% with the closest competitor being GED with average rate savings of 0.68\%.
Out of the remaining motion models, only 3DT is able to obtain rate savings on average, whereas all other motion models lead to increases in rate.
Comparing MPA to GED, it is visible that GED is able to outperform MPA on \textit{Landing2} and \textit{BranCastle2}.
While on \textit{Landing2}, GED achieves notable further rate savings over MPA, for \textit{BranCastle2}, the rate savings obtained by MPA and GED are close.
On the other hand, on the sequences \textit{SkateboardInLot}, \textit{Balboa}, and \textit{Broadway}, MPA is able to achieve significantly higher rate savings than GED showcasing its broad applicability for diverse motion characteristics.

\begin{table}
  \centering
  \renewcommand{\arraystretch}{1.1}
  \caption{Comparison of BD-Rate with respect to VTM-14.2 for different 360-degree motion models based on WS-PSNR. To allow a fair comparison, MPA-MVP=OFF for MPA. Negative values (black) represent actual rate savings, positive values (red) represent increases in rate. Bold entries mark the hightest rate savings among the tested motion models.}\label{table:models-table}
  \begin{tabular}{l|r|r|r|r|r|r}
& MPA & \makecell[tc]{3DT\\\cite{Li2017}\\\cite{Li2019}} & \makecell[tc]{TAN\\\cite{DeSimone2017}} & \makecell[tc]{ROT\\\cite{Vishwanath2017}\\\cite{Vishwanath2018a}} & \makecell[tc]{GED\\\cite{Vishwanath2018}} & \makecell[tc]{TAN+\\ROT\\\cite{Marie2021}}\\ 
\hline 
SkateboardInLot & \textbf{\textcolor{black}{-1.06}} & \textcolor{red}{+0.13} & \textcolor{red}{+0.78} & \textcolor{red}{+0.79} & \textcolor{black}{-0.13} & \textcolor{red}{+0.88} \\ 
ChairliftRide & \textbf{\textcolor{black}{-1.22}} & \textcolor{black}{-0.16} & \textcolor{red}{+0.10} & \textcolor{red}{+0.42} & \textcolor{black}{-1.19} & \textcolor{red}{+0.13} \\ 
KiteFlite & \textcolor{red}{+0.03} & \textcolor{red}{+0.08} & \textcolor{red}{+0.06} & \textbf{\textcolor{black}{-0.02}} & \textcolor{red}{+0.05} & \textcolor{red}{+0.29} \\ 
Harbor & \textcolor{red}{+0.27} & \textcolor{red}{+0.16} & \textcolor{red}{+0.31} & \textcolor{red}{+0.07} & \textcolor{red}{+0.13} & \textcolor{red}{+0.15} \\ 
Trolley & \textcolor{red}{+0.04} & \textcolor{red}{+0.16} & \textcolor{black}{-0.02} & \textcolor{red}{+0.23} & \textcolor{red}{+0.15} & \textbf{\textcolor{black}{-0.04}} \\ 
GasLamp & \textcolor{black}{-0.08} & \textcolor{black}{-0.10} & \textbf{\textcolor{black}{-0.74}} & \textcolor{black}{-0.47} & \textcolor{black}{-0.57} & \textcolor{black}{-0.64} \\ 
Balboa & \textbf{\textcolor{black}{-1.87}} & \textcolor{black}{-0.07} & \textcolor{red}{+0.59} & \textcolor{red}{+0.71} & \textcolor{black}{-0.53} & \textcolor{red}{+0.59} \\ 
Broadway & \textbf{\textcolor{black}{-1.64}} & \textcolor{red}{+0.11} & \textcolor{red}{+0.78} & \textcolor{red}{+0.50} & \textcolor{black}{-0.61} & \textcolor{red}{+0.52} \\ 
Landing2 & \textcolor{black}{-2.92} & \textcolor{black}{-1.03} & \textcolor{black}{-1.72} & \textcolor{red}{+0.25} & \textbf{\textcolor{black}{-3.50}} & \textcolor{black}{-0.99} \\ 
BranCastle2 & \textcolor{black}{-0.58} & \textcolor{black}{-0.22} & \textcolor{red}{+0.03} & \textcolor{red}{+0.18} & \textbf{\textcolor{black}{-0.62}} & \textcolor{red}{+0.15} \\ 
\hline 
Average & \textbf{-0.90} & \textcolor{black}{-0.09} & \textcolor{red}{+0.02} & \textcolor{red}{+0.27} & \textcolor{black}{-0.68} & \textcolor{red}{+0.10} \\ 
\end{tabular}
\end{table}

\subsection{Tool Utilization}

\begin{figure*}[t!]
  \centering
  \setlength{\framewidth}{0.48\linewidth}
  \subfloat[\textit{Landing2}, Frame 14, QP22, MPA-MVP=OFF\label{fig:mpa-cus:QP22-nomvp}]{%
  \includegraphics[width=\framewidth]{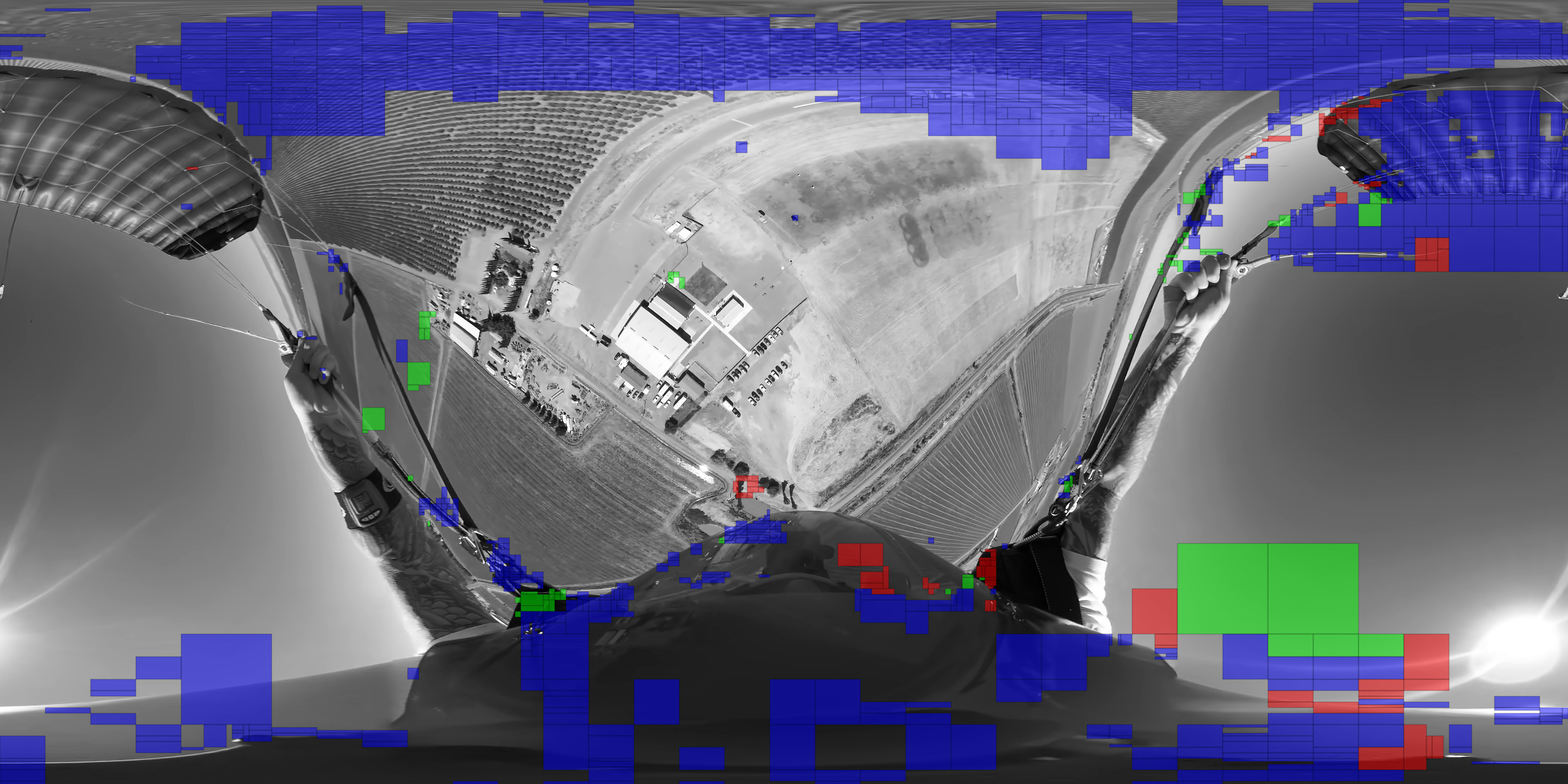}}
  \hfill
  \subfloat[\textit{Landing2}, Frame 14, QP22, MPA-MVP=ON\label{fig:mpa-cus:QP22-mvp}]{%
  \includegraphics[width=\framewidth]{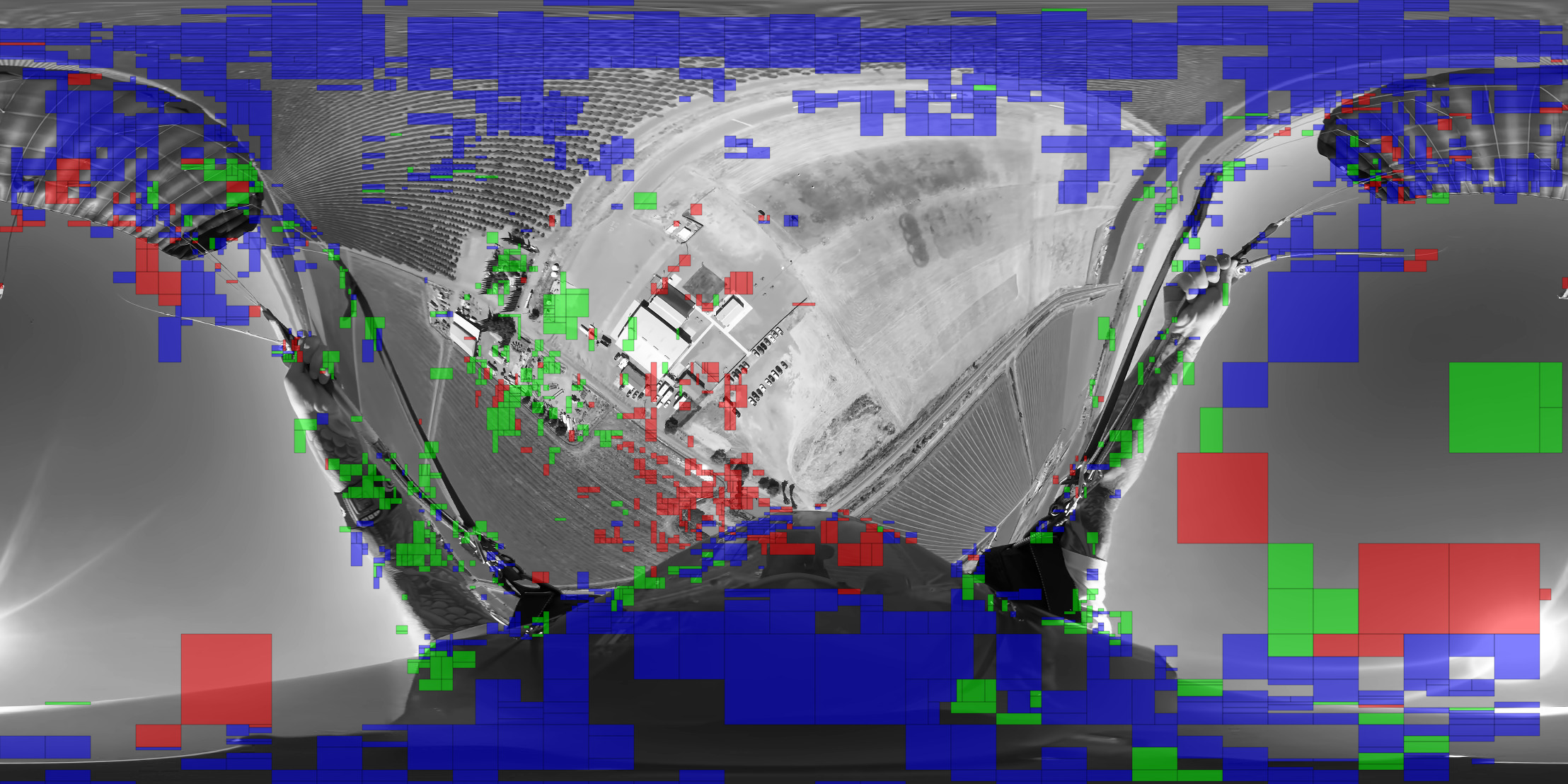}}
  \\
  \subfloat[\textit{ChairliftRide}, Frame 20, QP27, MPA-MVP=OFF\label{fig:mpa-cus:QP27-nomvp}]{%
  \includegraphics[width=\framewidth]{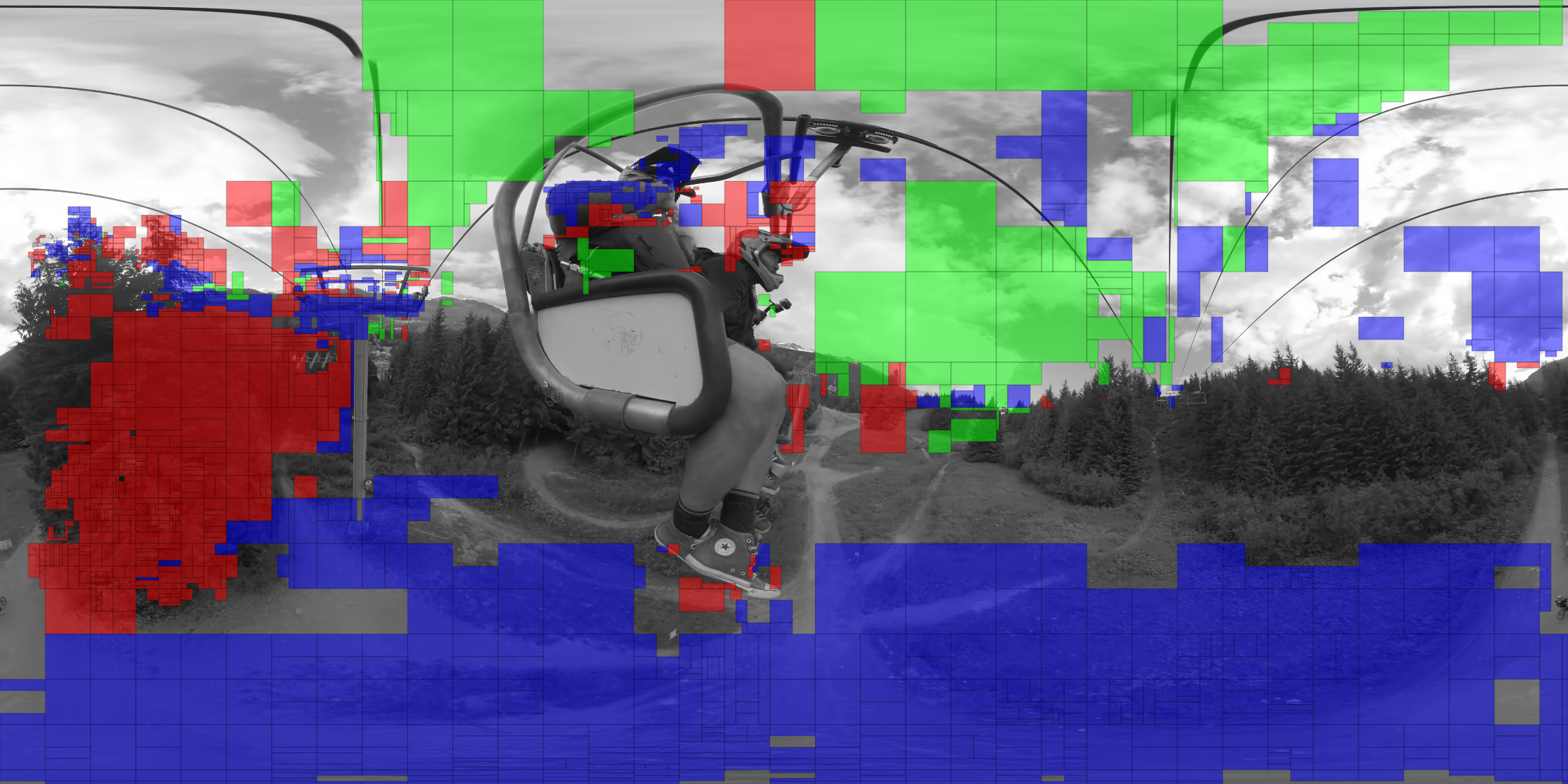}}
  \hfill
  \subfloat[\textit{ChairliftRide}, Frame 20, QP27, MPA-MVP=ON\label{fig:mpa-cus:QP27-mvp}]{%
  \includegraphics[width=\framewidth]{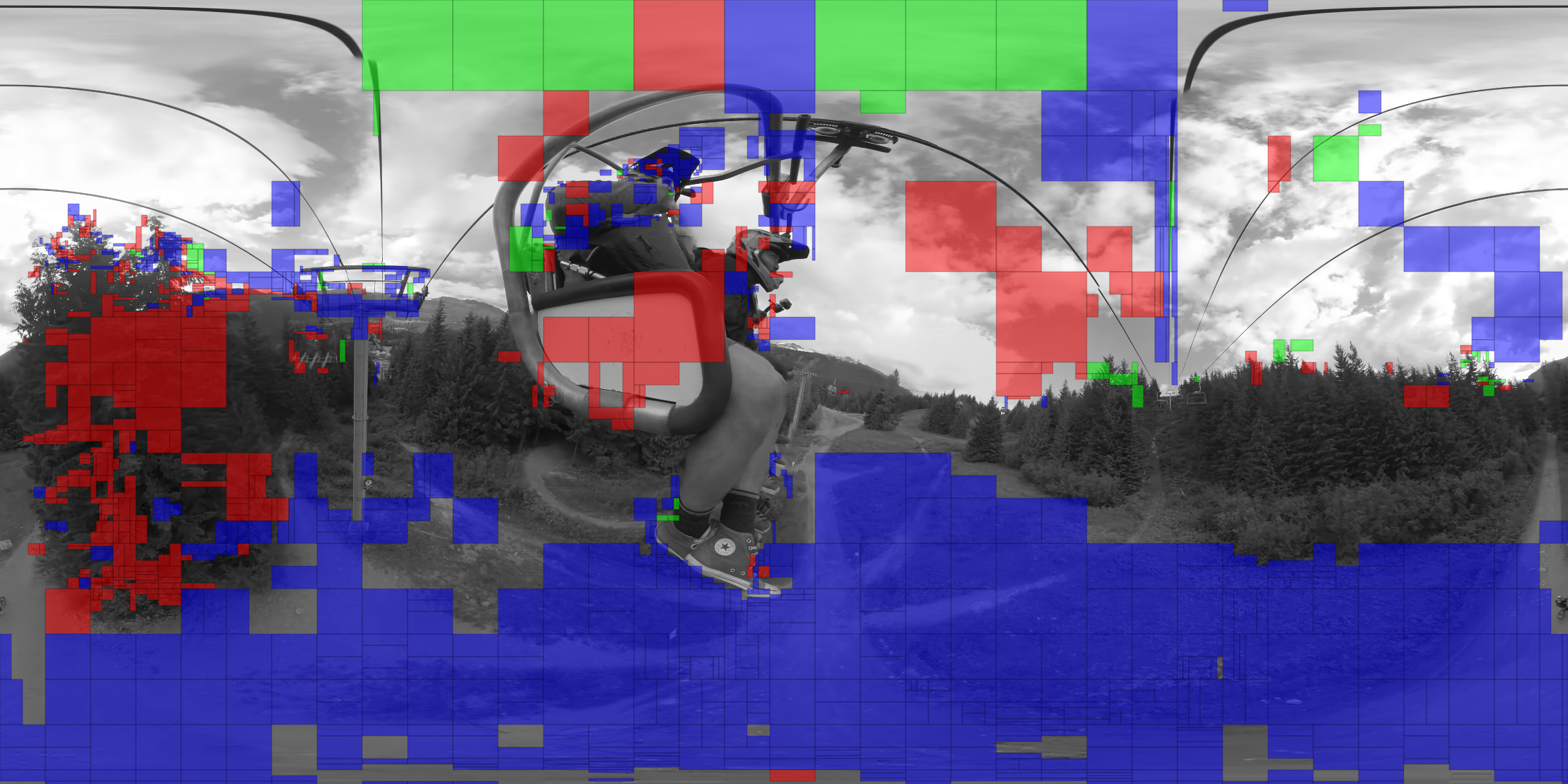}}
  \caption{MPA-CUs with motion planes color-coded as front/back (red), left/right (green), bottom/top (blue).}
  \label{fig:mpa-cus}
\end{figure*}

Table~\ref{table:utilization-table} displays the MPA-utilization as the average percentage of pixels predicted using the proposed MPA tool for each sequence for MPA-MVP=OFF and ON.
The average utilization of MPA lies at 4.95\% for MPA-MVP=OFF and at 5.76\% for MPA-MVP=ON.
While most sequences show an even higher MPA-utilization of more than 6\% and 7\% for MPA-MVP=OFF and ON, respectively, the four sequences \textit{KiteFlite}, \textit{Harbor}, \textit{Trolley} and \textit{Gaslamp} lie considerable below the average.
As stated before, these sequences inhibit no global camera motion and only little, locally constrained object motion.
The usage of MPA is therefore less prominent on these sequences as most CUs do not require an actual motion compensation at all.

\begin{table}
  \centering
  \renewcommand{\arraystretch}{1.1}
  \caption{Percentage of pixels predicted using MPA in MPA-VVC averaged over all frames.}\label{table:utilization-table}
  \begin{tabular}{l|r|r}
\multicolumn{1}{r|}{MPA-MVP} & OFF & ON \\ 
\hline 
SkateboardInLot & 6.86\% & 7.94\% \\ 
ChairliftRide & 10.38\% & 10.91\% \\ 
KiteFlite & 1.31\% & 1.39\% \\ 
Harbor & 1.00\% & 0.84\% \\ 
Trolley & 0.73\% & 0.65\% \\ 
GasLamp & 0.55\% & 0.60\% \\ 
Balboa & 7.30\% & 8.29\% \\ 
Broadway & 7.54\% & 8.80\% \\ 
Landing2 & 6.22\% & 7.28\% \\ 
BranCastle2 & 7.56\% & 10.94\% \\ 
\hline 
Average & 4.95\% & 5.76\% \\ 
\end{tabular}
\end{table}


Fig.~\ref{fig:mpa-cus} visualizes the motion plane maps, i.e., the motion plane decisions, for individual frames and QP values in the sequences \textit{Landing2} (Frame 14, QP=22) and \textit{ChairliftRide} (Frame 20, QP=27) for MPA-MVP=ON and OFF.
The motion plane decisions for each CU are color-coded as front/back (red), left/right (green) and bottom/top (blue).
Transparent areas denote that another inter prediction tool or intra prediction is used.

In the motion plane maps for the sequence \textit{Landing2} in Fig.~\ref{fig:mpa-cus:QP22-nomvp} and (b), the bottom/top motion plane is chosen especially in the highly distorted areas close to the north and south poles.
The introduction of MPA-MVP in Fig.~\ref{fig:mpa-cus:QP22-mvp} shows clear benefits as the utilization of MPA towards the south pole increases notably compared to MPA-MVP=OFF in Fig.~\ref{fig:mpa-cus:QP22-nomvp} and a lot of small, isolated CUs close the the center of the frame now employ MPA that previously used another tool.
As the motion planes of MPA are not aligned with the motion surfaces for this sequence due to rich global camera translation and rotation, these observations show that MPA is able to better replicate the distortions occuring in 360-degree video than the classical translational and affine motion models even in suboptimal scenarios.

\begin{figure*}[t]
\setlength\figurewidth{0.5\linewidth}
\setlength\figureheight{.3\linewidth}
\centering
{
  \footnotesize%
  \subfloat[Decoder\label{fig:complexity:decoder}]{%
\begin{tikzpicture}

\definecolor{color0}{rgb}{0.881862745098039,0.505392156862745,0.173039215686275}
\definecolor{color1}{rgb}{0.75343137254902,0.238725490196078,0.241666666666667}
\definecolor{color2}{rgb}{0.194607843137255,0.453431372549019,0.632843137254902}
\definecolor{color3}{rgb}{0.578431372549019,0.446078431372549,0.699019607843137}
\definecolor{color4}{rgb}{0.517156862745098,0.358333333333333,0.325980392156863}
\definecolor{color5}{rgb}{0.662254901960784,0.665196078431373,0.209313725490196}
\definecolor{color6}{rgb}{0.837254901960784,0.519607843137255,0.740196078431372}

\begin{axis}[
height=\figureheight,
minor xtick={},
minor ytick={},
tick align=outside,
tick pos=left,
ticklabel style={font=\scriptsize},
width=\figurewidth,
x grid style={white!69.0196078431373!black},
xlabel={Motion model},
xmajorgrids,
xmin=-0.5, xmax=6.5,
xtick style={color=black},
xtick={0,1,2,3,4,5,6},
xticklabel style={align=center},
xticklabels={MPA,MPA\\+MPA-MVP,3DT,TAN,ROT,GED,TAN+ROT},
y grid style={white!69.0196078431373!black},
ymajorgrids,
ymin=-8.15766880398363, ymax=803.571181858218,
ytick style={color=black},
ytick={0,100,200,300,400,500,600,700,800}
]
\path [draw=white!24.7058823529412!black, fill=color0]
(axis cs:-0.4,145.071912256065)
--(axis cs:0.4,145.071912256065)
--(axis cs:0.4,309.362166879553)
--(axis cs:-0.4,309.362166879553)
--(axis cs:-0.4,145.071912256065)
--cycle;
\path [draw=white!24.7058823529412!black, fill=color1]
(axis cs:0.6,152.796837236067)
--(axis cs:1.4,152.796837236067)
--(axis cs:1.4,302.09260292839)
--(axis cs:0.6,302.09260292839)
--(axis cs:0.6,152.796837236067)
--cycle;
\path [draw=white!24.7058823529412!black, fill=color2]
(axis cs:1.6,105.3543005171)
--(axis cs:2.4,105.3543005171)
--(axis cs:2.4,174.571033623368)
--(axis cs:1.6,174.571033623368)
--(axis cs:1.6,105.3543005171)
--cycle;
\path [draw=white!24.7058823529412!black, fill=color3]
(axis cs:2.6,129.869292902388)
--(axis cs:3.4,129.869292902388)
--(axis cs:3.4,246.98180022533)
--(axis cs:2.6,246.98180022533)
--(axis cs:2.6,129.869292902388)
--cycle;
\path [draw=white!24.7058823529412!black, fill=color4]
(axis cs:3.6,103.869015819786)
--(axis cs:4.4,103.869015819786)
--(axis cs:4.4,153.910595048446)
--(axis cs:3.6,153.910595048446)
--(axis cs:3.6,103.869015819786)
--cycle;
\path [draw=white!24.7058823529412!black, fill=color5]
(axis cs:4.6,148.150019533991)
--(axis cs:5.4,148.150019533991)
--(axis cs:5.4,437.315042148842)
--(axis cs:4.6,437.315042148842)
--(axis cs:4.6,148.150019533991)
--cycle;
\path [draw=white!24.7058823529412!black, fill=color6]
(axis cs:5.6,110.544021226915)
--(axis cs:6.4,110.544021226915)
--(axis cs:6.4,195.06348943153)
--(axis cs:5.6,195.06348943153)
--(axis cs:5.6,110.544021226915)
--cycle;
\addplot [white!24.7058823529412!black]
table {%
0 145.071912256065
0 45.771023454112
};
\addplot [white!24.7058823529412!black]
table {%
0 309.362166879553
0 472.754514727084
};
\addplot [white!24.7058823529412!black]
table {%
-0.2 45.771023454112
0.2 45.771023454112
};
\addplot [white!24.7058823529412!black]
table {%
-0.2 472.754514727084
0.2 472.754514727084
};
\addplot [white!24.7058823529412!black]
table {%
1 152.796837236067
1 82.5324712562532
};
\addplot [white!24.7058823529412!black]
table {%
1 302.09260292839
1 467.989004492074
};
\addplot [white!24.7058823529412!black]
table {%
0.8 82.5324712562532
1.2 82.5324712562532
};
\addplot [white!24.7058823529412!black]
table {%
0.8 467.989004492074
1.2 467.989004492074
};
\addplot [white!24.7058823529412!black]
table {%
2 105.3543005171
2 36.1751339267181
};
\addplot [white!24.7058823529412!black]
table {%
2 174.571033623368
2 244.242193588124
};
\addplot [white!24.7058823529412!black]
table {%
1.8 36.1751339267181
2.2 36.1751339267181
};
\addplot [white!24.7058823529412!black]
table {%
1.8 244.242193588124
2.2 244.242193588124
};
\addplot [black, mark=diamond*, mark size=2.5, mark options={solid,fill=white!24.7058823529412!black}, only marks]
table {%
2 355.514756430047
};
\addplot [white!24.7058823529412!black]
table {%
3 129.869292902388
3 44.3372856608307
};
\addplot [white!24.7058823529412!black]
table {%
3 246.98180022533
3 361.355371260534
};
\addplot [white!24.7058823529412!black]
table {%
2.8 44.3372856608307
3.2 44.3372856608307
};
\addplot [white!24.7058823529412!black]
table {%
2.8 361.355371260534
3.2 361.355371260534
};
\addplot [black, mark=diamond*, mark size=2.5, mark options={solid,fill=white!24.7058823529412!black}, only marks]
table {%
3 457.384116440438
};
\addplot [white!24.7058823529412!black]
table {%
4 103.869015819786
4 45.203300269452
};
\addplot [white!24.7058823529412!black]
table {%
4 153.910595048446
4 219.765307361338
};
\addplot [white!24.7058823529412!black]
table {%
3.8 45.203300269452
4.2 45.203300269452
};
\addplot [white!24.7058823529412!black]
table {%
3.8 219.765307361338
4.2 219.765307361338
};
\addplot [black, mark=diamond*, mark size=2.5, mark options={solid,fill=white!24.7058823529412!black}, only marks]
table {%
4 229.965197993239
4 254.904201663457
4 262.51385814806
};
\addplot [white!24.7058823529412!black]
table {%
5 148.150019533991
5 42.7638402196067
};
\addplot [white!24.7058823529412!black]
table {%
5 437.315042148842
5 766.674415919027
};
\addplot [white!24.7058823529412!black]
table {%
4.8 42.7638402196067
5.2 42.7638402196067
};
\addplot [white!24.7058823529412!black]
table {%
4.8 766.674415919027
5.2 766.674415919027
};
\addplot [white!24.7058823529412!black]
table {%
6 110.544021226915
6 28.7390971352073
};
\addplot [white!24.7058823529412!black]
table {%
6 195.06348943153
6 305.752135396243
};
\addplot [white!24.7058823529412!black]
table {%
5.8 28.7390971352073
6.2 28.7390971352073
};
\addplot [white!24.7058823529412!black]
table {%
5.8 305.752135396243
6.2 305.752135396243
};
\addplot [black, mark=diamond*, mark size=2.5, mark options={solid,fill=white!24.7058823529412!black}, only marks]
table {%
6 322.1630847307
6 359.007799495379
6 344.092785872922
};
\addplot [white!24.7058823529412!black]
table {%
-0.4 213.838490711609
0.4 213.838490711609
};
\addplot [white!24.7058823529412!black]
table {%
0.6 199.25907643716
1.4 199.25907643716
};
\addplot [white!24.7058823529412!black]
table {%
1.6 138.048442131051
2.4 138.048442131051
};
\addplot [white!24.7058823529412!black]
table {%
2.6 187.133097627854
3.4 187.133097627854
};
\addplot [white!24.7058823529412!black]
table {%
3.6 125.23121337986
4.4 125.23121337986
};
\addplot [white!24.7058823529412!black]
table {%
4.6 217.525212658253
5.4 217.525212658253
};
\addplot [white!24.7058823529412!black]
table {%
5.6 144.873420570896
6.4 144.873420570896
};
\end{axis}

\end{tikzpicture}}%
  \hfill%
  \subfloat[Encoder\label{fig:complexity:encoder}]{%
\begin{tikzpicture}

\definecolor{color0}{rgb}{0.881862745098039,0.505392156862745,0.173039215686275}
\definecolor{color1}{rgb}{0.75343137254902,0.238725490196078,0.241666666666667}
\definecolor{color2}{rgb}{0.194607843137255,0.453431372549019,0.632843137254902}
\definecolor{color3}{rgb}{0.578431372549019,0.446078431372549,0.699019607843137}
\definecolor{color4}{rgb}{0.517156862745098,0.358333333333333,0.325980392156863}
\definecolor{color5}{rgb}{0.662254901960784,0.665196078431373,0.209313725490196}
\definecolor{color6}{rgb}{0.837254901960784,0.519607843137255,0.740196078431372}

\begin{axis}[
height=\figureheight,
minor xtick={},
minor ytick={},
tick align=outside,
tick pos=left,
ticklabel style={font=\scriptsize},
width=\figurewidth,
x grid style={white!69.0196078431373!black},
xlabel={Motion model},
xmajorgrids,
xmin=-0.5, xmax=6.5,
xtick style={color=black},
xtick={0,1,2,3,4,5,6},
xticklabel style={align=center},
xticklabels={MPA,MPA\\+MPA-MVP,3DT,TAN,ROT,GED,TAN+ROT},
y grid style={white!69.0196078431373!black},
ylabel={Relative complexity in \%},
ymajorgrids,
ymin=0, ymax=3654.30054363213,
ytick style={color=black},
ytick={0,500,1000,1500,2000,2500,3000,3500}
]
\path [draw=black, fill=color0, postaction={pattern=north east lines}]
(axis cs:-0.4,500.420827203988)
--(axis cs:0.4,500.420827203988)
--(axis cs:0.4,1224.07359600754)
--(axis cs:-0.4,1224.07359600754)
--(axis cs:-0.4,500.420827203988)
--cycle;
\path [draw=black, fill=color1, postaction={pattern=north east lines}]
(axis cs:0.6,499.010368314891)
--(axis cs:1.4,499.010368314891)
--(axis cs:1.4,1249.95044702196)
--(axis cs:0.6,1249.95044702196)
--(axis cs:0.6,499.010368314891)
--cycle;
\path [draw=white!24.7058823529412!black, fill=color0]
(axis cs:-0.4,2257.23024975539)
--(axis cs:0.4,2257.23024975539)
--(axis cs:0.4,2809.58363272846)
--(axis cs:-0.4,2809.58363272846)
--(axis cs:-0.4,2257.23024975539)
--cycle;
\path [draw=white!24.7058823529412!black, fill=color1]
(axis cs:0.6,2385.02652764452)
--(axis cs:1.4,2385.02652764452)
--(axis cs:1.4,2964.61238119337)
--(axis cs:0.6,2964.61238119337)
--(axis cs:0.6,2385.02652764452)
--cycle;
\path [draw=white!24.7058823529412!black, fill=color2]
(axis cs:1.6,497.21159917633)
--(axis cs:2.4,497.21159917633)
--(axis cs:2.4,560.697696402523)
--(axis cs:1.6,560.697696402523)
--(axis cs:1.6,497.21159917633)
--cycle;
\path [draw=white!24.7058823529412!black, fill=color3]
(axis cs:2.6,792.981396725002)
--(axis cs:3.4,792.981396725002)
--(axis cs:3.4,959.627243033229)
--(axis cs:2.6,959.627243033229)
--(axis cs:2.6,792.981396725002)
--cycle;
\path [draw=white!24.7058823529412!black, fill=color4]
(axis cs:3.6,557.767374796645)
--(axis cs:4.4,557.767374796645)
--(axis cs:4.4,645.199668829214)
--(axis cs:3.6,645.199668829214)
--(axis cs:3.6,557.767374796645)
--cycle;
\path [draw=white!24.7058823529412!black, fill=color5]
(axis cs:4.6,698.305589432146)
--(axis cs:5.4,698.305589432146)
--(axis cs:5.4,930.916953389361)
--(axis cs:4.6,930.916953389361)
--(axis cs:4.6,698.305589432146)
--cycle;
\path [draw=white!24.7058823529412!black, fill=color6]
(axis cs:5.6,1526.18826399449)
--(axis cs:6.4,1526.18826399449)
--(axis cs:6.4,1933.62004078782)
--(axis cs:5.6,1933.62004078782)
--(axis cs:5.6,1526.18826399449)
--cycle;
\addplot [black, dash pattern=on 2.1pt off 1pt]
table {%
0 500.420827203988
0 175.972313540409
};
\addplot [black, dash pattern=on 2.1pt off 1pt]
table {%
0 1224.07359600754
0 1828.57627981074
};
\addplot [black, dash pattern=on 2.1pt off 1pt]
table {%
-0.2 175.972313540409
0.2 175.972313540409
};
\addplot [black, dash pattern=on 2.1pt off 1pt]
table {%
-0.2 1828.57627981074
0.2 1828.57627981074
};
\addplot [black, dash pattern=on 2.1pt off 1pt]
table {%
1 499.010368314891
1 179.026499513759
};
\addplot [black, dash pattern=on 2.1pt off 1pt]
table {%
1 1249.95044702196
1 2084.18613627444
};
\addplot [black, dash pattern=on 2.1pt off 1pt]
table {%
0.8 179.026499513759
1.2 179.026499513759
};
\addplot [black, dash pattern=on 2.1pt off 1pt]
table {%
0.8 2084.18613627444
1.2 2084.18613627444
};
\addplot [white!24.7058823529412!black]
table {%
0 2257.23024975539
0 1880.45355656547
};
\addplot [white!24.7058823529412!black]
table {%
0 2809.58363272846
0 3488.66586600872
};
\addplot [white!24.7058823529412!black]
table {%
-0.2 1880.45355656547
0.2 1880.45355656547
};
\addplot [white!24.7058823529412!black]
table {%
-0.2 3488.66586600872
0.2 3488.66586600872
};
\addplot [white!24.7058823529412!black]
table {%
1 2385.02652764452
1 1715.50537584865
};
\addplot [white!24.7058823529412!black]
table {%
1 2964.61238119337
1 3477.49636999902
};
\addplot [white!24.7058823529412!black]
table {%
0.8 1715.50537584865
1.2 1715.50537584865
};
\addplot [white!24.7058823529412!black]
table {%
0.8 3477.49636999902
1.2 3477.49636999902
};
\addplot [white!24.7058823529412!black]
table {%
2 497.21159917633
2 421.742555030874
};
\addplot [white!24.7058823529412!black]
table {%
2 560.697696402523
2 643.753443809676
};
\addplot [white!24.7058823529412!black]
table {%
1.8 421.742555030874
2.2 421.742555030874
};
\addplot [white!24.7058823529412!black]
table {%
1.8 643.753443809676
2.2 643.753443809676
};
\addplot [white!24.7058823529412!black]
table {%
3 792.981396725002
3 656.350235170167
};
\addplot [white!24.7058823529412!black]
table {%
3 959.627243033229
3 1149.05270546341
};
\addplot [white!24.7058823529412!black]
table {%
2.8 656.350235170167
3.2 656.350235170167
};
\addplot [white!24.7058823529412!black]
table {%
2.8 1149.05270546341
3.2 1149.05270546341
};
\addplot [white!24.7058823529412!black]
table {%
4 557.767374796645
4 491.561856348826
};
\addplot [white!24.7058823529412!black]
table {%
4 645.199668829214
4 741.314415072407
};
\addplot [white!24.7058823529412!black]
table {%
3.8 491.561856348826
4.2 491.561856348826
};
\addplot [white!24.7058823529412!black]
table {%
3.8 741.314415072407
4.2 741.314415072407
};
\addplot [white!24.7058823529412!black]
table {%
5 698.305589432146
5 593.260863468669
};
\addplot [white!24.7058823529412!black]
table {%
5 930.916953389361
5 967.65701589281
};
\addplot [white!24.7058823529412!black]
table {%
4.8 593.260863468669
5.2 593.260863468669
};
\addplot [white!24.7058823529412!black]
table {%
4.8 967.65701589281
5.2 967.65701589281
};
\addplot [black, mark=diamond*, mark size=2.5, mark options={solid,fill=white!24.7058823529412!black}, only marks]
table {%
5 1748.14258549841
5 1927.87368987467
5 2062.4875687243
5 2209.32425829864
5 2327.15094816552
5 1681.66973013082
5 1701.9445789353
5 1661.64693583296
5 1397.16471014172
};
\addplot [white!24.7058823529412!black]
table {%
6 1526.18826399449
6 1277.64997750296
};
\addplot [white!24.7058823529412!black]
table {%
6 1933.62004078782
6 2250.59427486101
};
\addplot [white!24.7058823529412!black]
table {%
5.8 1277.64997750296
6.2 1277.64997750296
};
\addplot [white!24.7058823529412!black]
table {%
5.8 2250.59427486101
6.2 2250.59427486101
};
\addplot [black, mark=diamond*, mark size=2.5, mark options={solid,fill=white!24.7058823529412!black}, only marks]
table {%
6 2627.06104799416
};
\addplot [black]
table {%
-0.4 834.582276092337
0.4 834.582276092337
};
\addplot [black]
table {%
0.6 896.330600266606
1.4 896.330600266606
};
\addplot [white!24.7058823529412!black]
table {%
-0.4 2535.11579891919
0.4 2535.11579891919
};
\addplot [white!24.7058823529412!black]
table {%
0.6 2807.74225208575
1.4 2807.74225208575
};
\addplot [white!24.7058823529412!black]
table {%
1.6 526.842715440847
2.4 526.842715440847
};
\addplot [white!24.7058823529412!black]
table {%
2.6 876.375436938245
3.4 876.375436938245
};
\addplot [white!24.7058823529412!black]
table {%
3.6 602.803026274877
4.4 602.803026274877
};
\addplot [white!24.7058823529412!black]
table {%
4.6 792.852386569206
5.4 792.852386569206
};
\addplot [white!24.7058823529412!black]
table {%
5.6 1742.6594130109
6.4 1742.6594130109
};
\draw (axis cs:0.5,2300) node[
  scale=1.0,
  anchor=base,
  text=black,
  rotate=0.0
]{$\uparrow$};
\draw (axis cs:0.5,2050) node[
  scale=0.8,
  anchor=base,
  text=black,
  rotate=0.0
]{Default};
\draw (axis cs:0.5,1500) node[
  scale=0.8,
  anchor=base,
  text=black,
  rotate=0.0
]{Fast};
\draw (axis cs:0.5,1300) node[
  scale=0.6,
  anchor=base,
  text=black,
  rotate=0.0
]{Min CU: 2048};
\draw (axis cs:0.5,1050) node[
  scale=1.0,
  anchor=base,
  text=black,
  rotate=0.0
]{$\downarrow$};
\end{axis}

\end{tikzpicture}}%
}
\caption{Boxplots visualizing the relative (a) decoder and (b) encoder complexity of the different motion models with respect to VTM-14.2. For MPA, the decoder and encoder complexity is shown for MPA-MVP=OFF (MPA) and ON (MPA+MPA-MVP). Furthermore, (b) shows a fast encoder configuration for MPA and MPA+MPA-MVP (hatched), where MPA is only enabled for CUs with a minimum size of 2048 pixels.}
\label{fig:complexity}
\end{figure*}
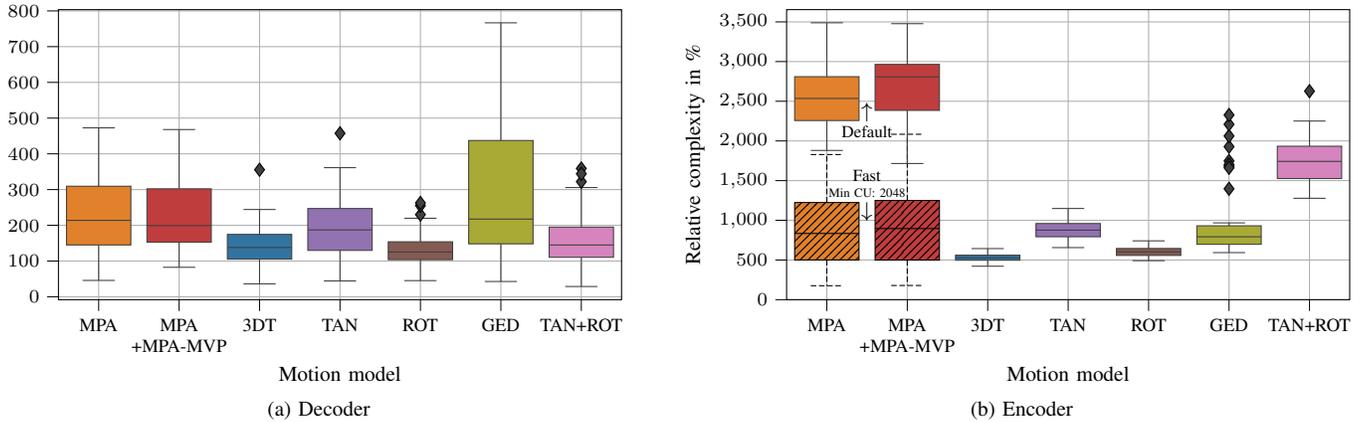

In the motion plane maps for the sequence \textit{ChairliftRide} in Fig.~\ref{fig:mpa-cus:QP27-nomvp} and (d), the motion surfaces in the sequence are better aligned with the available motion planes in MPA.
As such, the bottom/top motion plane is selected for most of the CUs on the ground surface, matching the intuitive choice.
Furthermore, the tree on the left of the image is coded using the front/back motion plane in many cases.
As this region of the image actually lies "behind" the camera in the spherical domain and the camera moves towards the right (along the cables), this choice can be intuitively expected, as well.
While the impact of MPA-MVP is less prominent here, the considerably reduced cost of switching the motion plane or motion model is still clearly visible as MPA with MPA-MVP=ON in Fig.~\ref{fig:mpa-cus:QP27-mvp} is more flexible in the selection of the best suitable parameters for each CU compared to MPA-MVP=OFF in Fig.~\ref{fig:mpa-cus:QP27-nomvp}.

\subsection{Complexity}

Fig.~\ref{fig:complexity:decoder} and (b) show boxplots representing the time complexity of the extended decoder and encoder for MPA and comparative 360-degree motion models with respect to the baseline VTM-14.2.

With a little more than 2-fold median complexity of MPA for MPA-MVP=OFF and slightly less than 2-fold complexity for MPA-MVP=ON with respect to VTM-14.2, the increase in decoder complexity as visible in Fig.~\ref{fig:complexity:decoder} is on a similar level as other 360-degree motion models.
However, as only GED showed convincing results in terms of rate-distortion performance (cf. Table~\ref{table:models-table}), a comparison of MPA to GED is most expressive.
With a median complexity of GED of a little more than 200\% with respect to VTM-14.2, the overall decoding complexities of MPA and GED are similar.
Nonetheless, the decoder for GED shows more extreme tendencies towards higher complexities with the third quartile lying at more than 400\% compared to roughly 300\% for MPA with MPA-MVP=OFF and ON.
Maximum complexities reach up to 800\% for GED and roughly 500\% for MPA.
Please note that further speed-ups can likely be achieved by runtime-optimized and/or hardware-specific implementations in the future.

However, with a median 25-fold complexity of MPA with MPA-MVP=OFF and 28-fold complexity of MPA with MPA-MVP=ON with respect to VTM-14.2, there is a significant increase in encoder complexity visible in Fig.~\ref{fig:complexity:encoder}.
Other motion models mostly lie in the range from 5-fold to 10-fold complexity, where GED has outliers going up to almost 25-fold complexity and TAN+ROT has a roughly 17-fold complexity as both motion model configurations need to be checked.
The increased encoder complexity for MPA with respect to other motion models is explicable by the motion estimation procedure needing to be performed three times for each of the available motion planes.
Comparing the complexity of MPA-MVP=OFF versus ON, it is visible that the introduction of MPA-MVP only adds little to the complexity of the encoder.
While the actual calculations required for MPA-MVP are negligible, MPA-MVP leads to more meaningful initializations of the motion vectors in the different motion models.
This leads to motion estimation with greedy search strategies like the so-called test zone search implemented in VTM-14.2 running for more iterations until it stops, explaining the slight increase in complexity.

\begin{table}
  \centering
  \renewcommand{\arraystretch}{1.1}
  \caption{Trade-off between obtained rate savings and relative encoder complexity for MPA with respect to VTM-14.2 based on WS-PSNR. Min CU defines the minimum number of pixels of a CU to allow the selection of the proposed MPA tool.}\label{table:sizeconstraint-table}
  \begin{tabular}{r||r|r||r|r}

\multicolumn{1}{r||}{MPA-MVP} & \multicolumn{2}{c||}{OFF} & \multicolumn{2}{c}{ON} \\ \hline
 \multicolumn{1}{r||}{Min CU} & BD-Rate & Complexity & BD-Rate & Complexity \\
\hline
      0 &  -0.90\% &      2617\% &  -1.56\% &      2851\% \\
    256 &  -1.15\% &      2191\% &  -1.39\% &      1749\% \\
    512 &  -1.00\% &      1703\% &  -1.35\% &      1327\% \\
   1024 &  -0.92\% &       829\% &  -1.16\% &       874\% \\
   2048 &  -0.76\% &       552\% &  -1.00\% &       610\% \\
   4096 &  -0.72\% &       434\% &  -0.83\% &       506\% \\
\hline
\end{tabular}

\end{table}

Due to the high encoding complexity at hand, we looked into methods to speed up our proposed approach.
Table~\ref{table:sizeconstraint-table} shows results for a simple, yet effective, complexity reduction, where a minimum CU size is required in order to enable MPA.
As expected, a trade-off between achieved rate savings (based on WS-PSNR) and complexity is adjustable via the minimum CU size.
With a more than 4-fold reduction in complexity for a minimum CU size of 2048 pixels for MPA-MVP=ON compared to MPA without a CU size constraint, and retained rate savings of 1\% on average, this simple technique demonstrates how a reduction of the search space can considerably speed up the encoding procedure.
The hatched boxplots in Fig.~\ref{fig:complexity:encoder} show that with this fast configuration, the encoding complexity of MPA is in a similar range as comparative 360-degree motion models.
In future research, more sophisticated approaches and smart search strategies are therefore likely to achieve even higher speed-ups while keeping the good compression performance of MPA.

\section{Conclusion}\label{sec:conclusion}

In this paper, we proposed a novel motion-plane-adaptive inter prediction technique called MPA that allows inter prediction on different motion planes in 3D space.
To let a video codec switch efficiently between different motion planes and motion models, we proposed a technique called MPA-MVP to translate motion information between them.
Our proposed integration of MPA together with MPA-MVP in the state-of-the-art H.266/VVC video coding standard showed significant average rate savings of 1.72\% based on PSNR and 1.56\% based on WS-PSNR with respect to VTM-14.2 for the RA configuration, where the novel MPA-MVP contributes around 0.6\% of the obtained rate savings on average.
MPA showed to be among the most competitive 360-degree motion models with its closest competitor, i.e., the geodesic motion model~\cite{Vishwanath2018}, reaching average rate savings of 0.68\% based on WS-PSNR.

In future research, we plan to investigate possibilities to reduce the complexity of MPA for both the encoder and the decoder by investigating optimized implementations for motion-plane-adaptive motion modeling and suitable fast encoder-side decision procedures.
Furthermore, we plan to analyze the achievable rate savings if MPA is applied to other 360-degree projection formats.

\bibliographystyle{IEEEtran}
\bibliography{ms}

\end{document}